\documentclass[journal=jctcce,manuscript=article]{achemso} 
\SectionNumbersOn
\setkeys{acs}{doi = true}
\usepackage{chemformula} 
\usepackage[T1]{fontenc} 
\usepackage{float} 
\usepackage{hyperref}
\usepackage{slashbox}
\usepackage{diagbox}
\usepackage{subcaption}
\usepackage{ltablex}
\usepackage{caption, booktabs}
\usepackage{tabularx}
\usepackage{tabularray}
\usepackage{comment}
\usepackage{natbib}
\usepackage{rotating}
\usepackage{xr-hyper}
\usepackage{cleveref}
\usepackage{colortbl}

\newcommand*\mycommand[1]{\texttt{\emph{#1}}}
\newcommand{\comm}[1]{}

\author{Marcin Ma\'zdziarz}
\affiliation{Institute of Fundamental Technological Research Polish Academy of Sciences, Warsaw, Poland}
\email{mmazdz@ippt.pan.pl}

\title{Uncertainty of DFT calculated mechanical and structural properties of solids due to incompatibility of pseudopotentials and exchange–correlation functionals}

\abbreviations{DFT, PP, XC}
\keywords{DFT, pseudopotentials, exchange–correlation functionals}

\begin{document}


\comm{Some journals require a graphical entry for the Table of Contents.
This should be laid out ``print ready'' so that the sizing of the
text is correct.

Inside the \texttt{tocentry} environment, the font used is Helvetica
8\,pt, as required by \emph{Journal of the American Chemical
Society}.

The surrounding frame is 9\,cm by 3.5\,cm, which is the maximum
permitted for  \emph{Journal of the American Chemical Society}
graphical table of content entries. The box will not resize if the
content is too big: instead it will overflow the edge of the box.

This box and the associated title will always be printed on a
separate page at the end of the document.
}

\begin{abstract}
	
The demand for pseudopotentials constructed for a given exchange-correlation (XC) functional far exceeds the supply, necessitating the use of those commonly available. The number of XC functionals currently available is in the hundreds, if not thousands, and the majority of pseudopotentials have been generated for the LDA and PBE. The objective of this study is to identify the error in the determination of the mechanical and structural properties (lattice constant, cohesive energy, surface energy, elastic constants, and bulk modulus) of crystals calculated by DFT with such inconsistency. Additionally, the study aims to estimate the performance of popular XC functionals (LDA, PBE, PBEsol, and SCAN) for these calculations in a consistent manner.  
	
\comm{This is an example document for the \textsf{achemso} document
  class, intended for submissions to the American Chemical Society
  for publication. The class is based on the standard \LaTeXe\
  \textsf{report} file, and does not seek to reproduce the appearance
  of a published paper.

  This is an abstract for the \textsf{achemso} document class
  demonstration document.  An abstract is only allowed for certain
  manuscript types.  The selection of \texttt{journal} and
  \texttt{manuscript} will determine if an abstract is valid.  If
  not, the class will issue an appropriate error. }
\end{abstract}

\section{Introduction}
\label{sec:Int}

\comm{	
	https://punktoza.pl/
	
	jctcce=Journal of Chemical Theory and Computation https://pubs.acs.org/page/jctcce/about.html 140p IF 6.578 it
	
	Computer Physics Communications https://www.sciencedirect.com/journal/computer-physics-communications  140p IF 4.717 it
	
	npj Computational Materials https://www.nature.com/npjcompumats/ 200p IF 12.241 im it €2590
	
	Science Bulletin https://www.sciencedirect.com/journal/science-bulletin 140p IF 20.577 im
	
	 Journal of Computational Chemistry https://onlinelibrary.wiley.com/journal/1096987X?utm_medium=web&utm_source=wileysjf 100p
	 
Policzyc rozne srednie i porownac https://pl.wikipedia.org/wiki/Nier
}

A fairly common approach in the density functional theory (DFT) \cite{DFT-HK, DFT-KS} calculations is the use of pseudopotentials (PPs), i.e. we replace the atomic all-electron potential in such a manner that core states are excluded from the total electron configuration and valence electrons are described by pseudo-wavefunctions and thus we decrease the size of the basis, the number of electrons, and generally reduce the cost of computation. This approach is used both in codes that utilize plane-wave (PW) basis for calculations, e.g. \href{https://www.vasp.at/}{VASP} \cite{KRESSE199615}, \href{https://www.abinit.org/}{ABINIT} \cite{GONZE2016106, GONZE2020107042}, \href{https://www.quantum-espresso.org/}{QUANTUM ESPRESSO} (QE) \cite{Giannozzi_2009, Giannozzi_2017} and many others, as well as those that use atomic-like orbitals (AO) basis, e.g. \href{https://www.cp2k.org/}{CP2K} \cite{CP2K2020} or \href{https://departments.icmab.es/leem/siesta/}{SIESTA} \cite{Soler_2002}. Norm-conserving (NC), ultrasoft (US) and the projector augmented wave (PAW) approach are the most common forms of pseudopotentials used in modern DFT codes. From the point of view of their accuracy, i.e. agreement with the results from all-electron (AE) calculations, the latest implementations of these three approaches are absolutely comparable, see \cite{Lejaeghere2016}. {It is worth mentioning that the AE codes also differ in their implementations, and the difference in results between them measured by the $\Delta$ gauge is comparable to that which is the result of the pseudization scheme.} It is important that the pseudopotentials are constructed for one specific exchange-correlation (XC) functional. Although XC functional is formally accurate \cite{DFT-HK}, an approximation must be chosen. Probably in the most popular library of exchange–correlation functionals for density functional theory, i.e. \href{https://tddft.org/programs/libxc/}{LibXC} \cite{LIBXC2018}, is included several hundred functionals for the exchange, correlation, and the kinetic energy. As exchange and correlation functionals can be formally combined, this gives many thousands of possibilities. 

In this plethora of XC functionals, some have gained particular appreciation and popularity because of their desirable features, and so, for example, the local density approximation (LDA) \cite{BlochBemerkungZE1929, Perdew1992} is derived directly from the homogeneous electron gas (HEG) model, classical Perdew--Burke--Ernzerhof (PBE) generalised gradient approximation (GGA) \cite{Perdew1996} and modified Perdew--Burke--Ernzerhof GGA for solids (PBEsol) \cite{Perdew2008} are popular due to being non-empirical, having low cost and good performance in semi-local pseudopotentials family.
Hybrid functionals are the next class of XC functionals in DFT that incorporate a portion of exact exchange from Hartree–Fock theory with the rest of the XC energy from other sources. They have gained popularity due to the improved calculation of some molecular properties and band gaps in crystals \cite{Burke1996}, unfortunately are non-local and their computational cost is much higher than those of LDA and GGA. Probably the most popular in this class of functionals are PBE0 \cite{Burke1996}, B3LYP \cite{Becke1993} and HSE06 \cite{Heyd2003}.    
The newest class of functionals are meta-generalized gradient approximations (MGGA), which are generally modified GGA enriched with kinetic energy. The advantage of these functionals is that are semi-local and therefore computationally efficient. They significantly improve the determination of the band gap \cite{Borlido2019} and, as far as accuracy is concerned, can outperform that for GGA. Probably here the most popular are TPSS \cite{Tao2003, Perdew2004} and strongly constrained and appropriately normed (SCAN) \cite{Sun2015} functionals.

And for how many XC functionals are the pseudopotentials constructed/available? Typical practice during DFT calculations is to use generated pseudopotentials from either online databases or those attached to code sources. In the \href{http://www.pseudo-dojo.org/}{PseudoDojo} we can found NC \cite{VANSETTEN201839} and PAW \cite{JOLLET20141246} pseudopotentials dedicated to \href{https://www.abinit.org/}{ABINIT}, \href{https://www.quantum-espresso.org/}{QUANTUM ESPRESSO}, \href{https://departments.icmab.es/leem/siesta/}{SIESTA}, constructed for LDA, PBE and PBEsol XC functionals. 
A standard solid-state pseudopotentials (\href{https://www.materialscloud.org/discover/sssp/table/efficiency}{SSSP}) library optimized for precision or efficiency \cite{Prandini2018} provides combination of NC/US/PAW PPs generated by the various codes only for PBE and PBEsol XC functionals. \href{https://dalcorso.github.io/pslibrary/}{PSlibrary} \cite{DALCORSO2014337} is a library of US and PAW PPs constructed for LDA, PBE and PBEsol XC functionals. Both these libraries are dedicated to \href{https://www.quantum-espresso.org/}{QUANTUM ESPRESSO}. \href{https://www.vasp.at/}{VASP} provides its own \href{https://www.vasp.at/wiki/index.php/Available_PAW_potentials}{PAW} PPs but only constructed for LDA and PBE XC functionals.
\href{https://www.cp2k.org/}{CP2K} provides with its own code the Goedecker-Teter-Hutter (GTH) pseudopotentials \cite{Goedecker1996, Krack2005} constructed for LDA, PBE, BLYP \cite{Becke1988, Lee1988}, PBE0 and SCAN. 

Scouring the Internet it is possible to find some PPs generated for other XC functionals but only for certain elements e.g. \href{https://nninc.cnf.cornell.edu/dd_search.php?frmxcprox=GGA&frmxctype=TPSS&frmspclass=}{BLYP, TPSS}, \href{https://yaoyi92.github.io/scan-tm-pseudopotentials.html}{SCAN}\cite{Yao2017}.  

If no one else has generated the PP you need, you can try to generate it using available \href{https://www.quantum-espresso.org/other-resources/}{generators}. The overwhelming majority of them generate only PPs for LDA and GGA XC functionals. The author found only two codes that generate PPs for MGGA XC functionals. The first is the Atomic Pseudo-potentials Engine \cite{OLIVEIRA2008524} (\href{https://tddft.org/programs/APE/}{APE}. Unfortunately, the author failed to generate PP for TPSS and SCAN XC functional with it. It seems that APE has not been updated recently and does not want to work with newer versions of LibXC. The second code is \href{http://users.wfu.edu/natalie/papers/pwpaw/}{ATOMPAW} \cite{Holzwarth2022} generating PAW PPs for ABINIT and QE. Unfortunately, ABINIT has limited functionality for MGGA XC functionals, stresses are not programmed which makes optimization of structures somewhat tedious. Present QE does not support PAW PPs for MGGA XC functionals. It is worth adding, following the Authors of the VASP manual, that the generation of pseudopotential is a kind of black art, and during generation weird things may happen.    

There seems to be little research on the effect of PPs and XC functionals incompatibility on DFT results. The error in the band gap when standard LDA/PBE PPs are used inconsistently to perform electronic structure calculations using other XC functionals is about 0.1\,eV, resulting in absolute relative errors in the 5–10\% range \cite{Borlido2020}. However, the Authors suggest that if accurate numerical estimates of band gaps are required, consistent PPs calculations are strongly recommended. Other Authors confirm these observations \cite{Bartok2019} and even write that PPs-XC functionals incompatibility leads to significant
differences in the resulting electronic structure compared to all-electron calculations, and that this practice should be avoided. For the PBE0 and HSE06 hybrid functionals\cite{Yang2023}, the error in band gap determination due to the use of inconsistent PBE pseudopotentials was 1.4\% for the four chemical compounds. On the other hand, for the tested MgO and AlP, the equilibrium lattice parameters, bulk moduli, and atomization energies are slightly less affected, with maximum relative consistency errors of 0.1\% (0.002 \AA), 0.8\% (0.7\,GPa), and 0.7\% (28\,meV/atom), respectively. 
For the PBE and PBE0 functionals \cite{Yang2018}, comparative tests on the G2 data set and solids indicate that the systematic lattice constant error associated with the inconsistency of the PPs-XC functionals is on the order of 1\%.
The cohesive properties of a set of typical metallic, semiconducting, and insulating crystals were also analyzed with the inconsistent GGA approach where LDA rather than GGA PPs were employed, but without a statistical analysis of the results to draw any general conclusions\cite{Garcia1992, Fuchs1998}.

In the case of predicting covalent bond energies for chemically relevant systems, the inconsistent use of PPs has been found to be an often overlooked but potentially serious source of error in modern DFT calculations\cite{Rossomme2023}.

To the best of my knowledge, there is no such study that systematically analyzes the error in the determination of mechanical and structural properties (lattice constant, cohesive energy, surface energy, elastic constants, and bulk modulus) of crystals calculated by DFT due to the inconsistency of PPs and XC functionals. {The purpose of this paper is not to estimate the pseudization error, i.e. the error introduced by replacing the core electrons with a pseudopotential, but to estimate the inconsistency error of the PPs and XC functionals, and the performance evaluation of the XC functionals used is only a free add-on here.}    

\section{Methods and Algorithms}
\label{sec:MandA} 

A subset of nine cubic solids, comprising three metallic, three covalent, and three ionic structures, was selected for analysis from a known collection of 20 cubic solids\cite{Sun2011}{}: Na (cI2, Im$\overline{3}$m, 229), Li (cI2, Im$\overline{3}$m, 229), Rh (cF4, Fm$\overline{3}$m, 225), C (cF8, Fd$\overline{3}$m, 227), Si (cF8, Fd$\overline{3}$m, 227), Ge (cF8, Fd$\overline{3}$m, 227), NaCl (cF8, Fm$\overline{3}$m, 225), LiCl (cF8, Fm$\overline{3}$m, 225), LiF (cF8, Fm$\overline{3}$m, 225). From this collection come the experimental data used here as reference: lattice constant a$_0$\,(\AA), cohesive energy E$_\mathrm{coh}$\,(eV/atom), and  bulk modulus \textbf{B}\,(GPa). Obtaining reliable empirical low-temperature elastic constants is a significant challenge, with results exhibiting considerable scatter, potentially reaching 20\% \cite{HUNTINGTON1958213}. {In the present work, only cubic crystals are used; averaging elastic constants for crystals from different crystal systems would de facto mean comparing stiffness tensors of different patterns\cite{GOLESORKHTABAR20131861}. 
	{The experimental determination of surface energies, particularly with respect to specific facets, is a challenging and infrequent undertaking. Furthermore, the consistency of observed Wulff shapes across different studies is often compromised due to the susceptibility of high-energy facets to temperature fluctuations and the effects of impurities \cite{Xu2016}. Experimentally measured surface energies are typically the "average" surface energies of all possible surfaces present in the analyzed sample. Reproducing these effects in DFT calculations is significantly difficult.} 
	
	Bulk modulus (\textbf{B}) is calculated here using the elastic constants (C$_{11}$, C$_{12}$) \cite{GOLESORKHTABAR20131861} and not the equation of state as in the\cite{Sun2011}{}:
	\begin{equation}
		\mathrm{B} = \dfrac{\mathrm{C_{11}}+2\,\mathrm{C_{12}}}{3}. 
		\label{eq:B}
	\end{equation}
	
	To quantify the results, the absolute percentage error (APE) defined as follows was used:
	
	\begin{equation}
		\mathrm{APE_\mathrm{DFT}^\mathrm{PP/XC} = \left |\frac{Value_{DFT}^{PP/XC}-Value_{DFT}^{PP=XC/XC}}{Value_{DFT}^{PP=XC/XC}}\right|{100\%}, }
		\label{eq:mapepp}
	\end{equation}
	
	\begin{equation}
		\mathrm{APE_\mathrm{EXP}^\mathrm{PP/XC} = \left |\frac{Value_{DFT}^{PP/XC}-Value_{EXP}^{}}{Value_{EXP}^{}}\right|{100\%}},
		\label{eq:mapee}
	\end{equation}
	
	in $\mathrm{APE_{DFT}^{PP/XC}}$ reference data are those from DFT calculations (PP is consistent with XC), while in $\mathrm{APE_\mathrm{EXP}^\mathrm{PP/XC}}$ the reference data are those from the experiment. {The variables in the present DFT calculation are the XC functional used and the PP pseudopotential generated for a certain XC functional, so PP/XC here means the combination of PP and XC one used and PP=XC means coherent case.}
	
	The average of the individual APEs' will represent the mean absolute percentage error (MAPE). Using non-relative error measures such as mean error (ME) or mean absolute error (MAE) when the bulk in the set varies by a factor of 60 times is dubious. \comm{The use of non-relative measures of error when the bulk modulus in a set varies by a factor of 60 is dubious.}
	
	{Another method applied to benchmark the compatibility of different codes, formulations, pseudizations, libraries, implementations of the DFT is to use the so-called $\Delta$ gauge\cite{Lejaeghere2016, Lejaeghere2014} , which measures the difference between the equations of state (EOS) for two different calculations.
		Comparing the results for every element $i$, the quantity $\Delta_i$ is defined as the root-mean-square difference between the EOS of computations $a$ and $b$ over a $\pm$6\% interval around
		the equilibrium volume $V_{0,i}$ :
		
		\begin{equation}
			{\Delta_\mathrm{i}(a,b) = \sqrt{\int_{1.06V_{0,i}}^{0.94V_{0,i}} \dfrac{(E_{b,i}(V) - E_{a,i}(V))^2}{0.12V_{0,i}} \,dV} } .
			\label{eq:deltapp}
		\end{equation}

		Based on the lattice constants and bulk moduli already calculated, we determine here the 2nd-order Birch-Murnaghan EOS and compare the coherent (PP=XC) and incoherent (PP$\neq$XC) calculations. }
	
	All the \emph{ab initio} calculations in this study were carried out in the density functional theory (DFT) \cite{DFT-HK, DFT-KS} formalism and using pseudopotential (PP) approach. The choice of DFT codes was determined by the availability of pseudopotentials consistent with the XC functionals. Since three PPs (LDA, PBE, PBEsol) are considered for the plane-wave (PW) basis and two PPs (PBE, SCAN) for the atomic-like orbitals (AO) basis, and since we analyze the mutual cross-checking of pseudopotentials {(n$^2$ combinations)}, we have a total of 117 cases (9$\times$3$\times$3 + 9$\times$2$\times$2) for nine analyzed solids.

\subsection{LDA, PBE, PBEsol - plane-wave (PW) basis}
\label{sec:PW} 

For the case of pseudopotentials constructed for LDA\cite{BlochBemerkungZE1929, Perdew1992}, PBE\cite{Perdew1996} and PBEsol\cite{Perdew2008} XC functionals the first-principle non-polarized spin-restricted DFT\cite{DFT-HK, DFT-KS} calculations within the pseudopotential plane-wave approximation (PP-PW) implemented in \href{https://www.abinit.org/}{ABINIT} \cite{GONZE2016106, GONZE2020107042} (v9.6.2) code were utilized in this study.

The precision of all the calculations was tuned by automatically setting the variables at \emph{accuracy} level 4 (\emph{accuracy}=4 corresponds to the default tuning of ABINIT). The electron configuration and \emph{cut-off} energy used in the calculations is in line with the hints for the given optimized norm-conserving Vanderbilt pseudopotentials \cite{Hamann2013} (ONCVPP) posted on database \href{http://www.pseudo-dojo.org/}{PseudoDojo}. K-PoinTs grids were generated with \emph{kptrlen}=50.0 (grids that give a length of smallest vector {larger} than \emph{kptrlen}). Metallic occupation of levels with the Gaussian smearing scheme and \emph{tsmear} (Ha)=0.02 was applied in all the ABINIT computations. All the generated structures were firstly relaxed by using the BFGS minimization scheme with full optimization of cell geometry and atomic coordinates. Maximal stress tolerance (GPa) was set to 1$\times$10$^{-4}$. Next, the theoretical ground state, T=0\,K, elastic constants $\mathrm{C_{ij}}$ of all the relaxed earlier structures were computed by means of the metric tensor formulation of strain in density functional perturbation theory (DFPT) \cite{Hamann2005}. The cohesive energy E$_\mathrm{coh}$\,(eV/atom) was derived as the difference in energy of a given relaxed earlier crystal and its individual atoms placed in a sufficiently large box \cite{Mazdziarz2016878}.
{The surface energy per unit area \textbf{$\gamma$} is typically defined as the energy required to separate the bulk material by a specific plane \cite{Parga2023}. The total energy of a slab of the material with a thickness of N layers is subtracted from the bulk energy. Unfortunately, the calculated surface energy oscillates regardless of the number of layers and does not converge to a specific value \cite{Fiorentini1996}.
	However, experience shows that the choice of a thickness of 4 layers and a similar thickness of vacuum gives good results and allows to ignore the small effect of surface relaxation, see ABINIT's tutorial. A surface with an orientation of (100) was selected for the calculations.}
Of course, there are many other properties that can be determined using DFT, but the quantities used here are readily available from the total energy calculations\cite{Lejaeghere2014}.}

\subsection{PBE, SCAN - atomic-like orbital (AO) basis}
\label{sec:AO} 

For the case of pseudopotentials constructed for PBE\cite{Perdew1996} and SCAN\cite{Sun2015} XC functionals the first-principle non-polarized spin-restricted DFT calculations were performed using the QUICKSTEP module of \href{https://www.cp2k.org/}{CP2K}\cite{CP2K2020} software (v2023.1) via the combined Gaussian and plane wave (GPW) method with the DZVP-MOLOPT-GTH\cite{VandeVondele2007} basis sets and the Goedecker-Teter-Hutter (GTH) pseudopotentials \cite{Goedecker1996}. 
The CP2K MULTIGRID method was used with four grids of increasing fineness, with the finest grid having an 300\,Ry energy \emph{cut-off} for the auxiliary grid and the reference level for mapping the Gaussians to the multigrid being set to 60\,Ry.

The Fermi-Dirac smearing with an electronic temperature of 300\,K, the Broyden mixing with target accuracy for the self-consistent field (SCF) convergence set to 10$^{-7}$ and the Monkhorst-pack 4$\times$4$\times$4 K-PoinTs grids scheme for the Brillouin zone sampling were used in the CP2K computations. During geometry relaxation by BFGS method pressure tolerance was set to 0.01\,bar.

Because the elastic constants cannot be calculated in CP2K using DFPT {, i.e., solving the 2$^{\text{nd}}$ order Sternheimer equation,} they were determined here by the stress-strain method. Since we have here crystals with cubic symmetry then a single deformation is sufficient to determine the three independent elastic constants, i.e. $\boldsymbol{\eta}=(\eta, 0, 0, \eta, 0, 0)$ and $\eta = 0.005$ \cite{GOLESORKHTABAR20131861}. It is worth remembering that, unfortunately, the results for this method are sensitive to the size of the perturbation, the method itself can introduce an uncertainty of up to several{\,}\%. The cohesive energy E$_\mathrm{coh}$\,(eV/atom) and the surface energy per unit area \textbf{$\gamma$} were calculated similarly to the plane-wave (PW) basis and ABINIT.

\section{Results and discussion}
\label{sec:RandD} 

All the detailed results (lattice constant a$_0$\,(\AA), cohesive energy E$_\mathrm{coh}$\,(eV/atom), surface energy \textbf{$\gamma$}\,(J/m$^2$), elastic constants C$_\mathrm{ij}$\,(GPa), and  bulk modulus \textbf{B}\,(GPa)) as well as determined APEs defined in equations~\ref{eq:mapepp}~and~\ref{eq:mapee} and {$\Delta_i$ gauge\,(meV/atom) defined in equation~\ref{eq:deltapp}} of the DFT calculations are gathered in Tables~\ref{tab:Na} - \ref{tab:LiF} in the {Supplementary material}\ref{sec:SI}. To verify the validity of the calculations performed, let us first compare the mean absolute percentage error ${\mathrm{MAPE_{EXP}}^\mathrm{PP/XC}}$\,(Eq.\ref{eq:mapee}) obtained here with a plane-wave basis (pseudopotential consistent with XC functional) and those in the paper with a fairly similar methodology\cite{Sun2011}(VASP, PW and PAW PPs). Here it was obtained: ${\mathrm{MAPE_{a_0}}^\mathrm{LDA/LDA}}$=1.463, ${\mathrm{MAPE_{a_0}}^\mathrm{PBE/PBE}}$=1.295, ${\mathrm{MAPE_{E_{coh}}}^\mathrm{LDA/LDA}}$=9.568, ${\mathrm{MAPE_{E_{coh}}}^\mathrm{PBE/PBE}}$=5.711, ${\mathrm{MAPE_{\textbf{B}}}^\mathrm{LDA/LDA}}$=10.885, ${\mathrm{MAPE_{\textbf{B}}}^\mathrm{PBE/PBE}}$=13.279, see Tab.\ref{tab:meanexp}, whereas in the aforementioned paper\cite{Sun2011}: {${\mathrm{MAPE_{a_0}}^\mathrm{LDA/LDA}}$=1.73}, {${\mathrm{MAPE_{a_0}}^\mathrm{PBE/PBE}}$=1.29}, ${\mathrm{MAPE_{E_{coh}}}^\mathrm{LDA/LDA}}$=16.50, ${\mathrm{MAPE_{E_{coh}}}^\mathrm{PBE/PBE}}$=4.23, ${\mathrm{MAPE_{\textbf{B}}}^\mathrm{LDA/LDA}}$=10.77, ${\mathrm{MAPE_{\textbf{B}}}^\mathrm{PBE/PBE}}$=10.61. As can be seen, the differences are tolerably small, due to computational details. 

{Although we do not have equivalent empirical data here, we can compare the surface energies \textbf{$\gamma$} calculated using a similar methodology. The surface energies for the unrelaxed surface (100) for the LDA XC functional were 9.72\,(J/m$^2$) for C, 2.39\,(J/m$^2$) for Si, and 1.71\,(J/m$^2$) for Ge \cite{Stekolnikov2002}, while in the present work they were 9.590\,(J/m$^2$) for C, 2.226\,(J/m$^2$) for Si, and 1.742\,(J/m$^2$) for Ge, see Tabs.\ref{tab:C}-\ref{tab:Ge}. The surface energy for Rh for the unrelaxed surface (100) was calculated for different XC functionals (inconsistently unfortunately) and was obtained for LDA 3.04\,(J/m$^2$), for PBE 2.77\,(J/m$^2$), for PBEsol 2.97\,(J/m$^2$), and for SCAN 2.71\,(J/m$^2$) \cite{Abhirup2017}. Here it was obtained 3.274\,(J/m$^2$), 2.467\,(J/m$^2$), 2.752\,(J/m$^2$) and 2.786\,(J/m$^2$) respectively, see Tab.\ref{tab:Rh}. The consistency of the results again looks satisfactory. 	
	}
	Unfortunately for the atomic-like orbital basis and pseudopotential consistent with the SCAN XC functional, we do not have such comprehensive reference calculations for the cubic solids analyzed here, and we must use available rare partial results for a plane-wave basis\cite{Yao2017}. They obtained a bulk modulus \textbf{B} of 99.30\,(GPa) for diamond phase silicon and \textbf{B} of 73.38\,(GPa) for germanium, whereas here it was obtained \textbf{B}=96.437\,(GPa) for silicon, see Tab.\ref{tab:Si}, and \textbf{B}=73.688\,(GPa) for germanium, see Tab.\ref{tab:Ge}. Although the DFT calculations were carried out in different bases, the agreement of the calculated bulk moduli is very good. It is estimated that using different bases in DFT calculations, plane waves and atomic-like orbitals, can result in differences of up to 1.8\% in lattice constants and up to 12.4\% in bulk moduli\cite{Skylaris2007}.
	
	The main objective of this work is to estimate the error in DFT calculations for the determination of structural and mechanical properties of crystals when the generated pseudopotentials used are inconsistent with the XC functionals and the principal findings are summarized in Tab.\ref{tab:meanpp}. When PP is generated for LDA and XC functional is GGA or PP is GGA and XC functional is LDA, the mean absolute percentage error ${\mathrm{MAPE_{DFT}}^\mathrm{PP/XC}}$\,(Eq.\ref{eq:mapepp}) for a$_0$\,(\AA) is 0.204\% with a standard deviation of 0.183\%, for E$_\mathrm{coh}$\,(eV/atom) is 0.820(0.645)\%, for \textbf{$\gamma$}\,(J/m$^2$) is 2.031(2.730)\%, for elastic constants C$_\mathrm{ij}$\,(GPa) and bulk modulus \textbf{B}\,(GPa) is about 1(0.8)\%, and the average for all properties \textbf{$\overline{\sum}$} is 1.042(1.322)\%. \comm{0.877(0.779)\%} 
	
	If the XC functional is PBE/PBEsol and PP is inconsistent with it but also GGA then the ${\mathrm{MAPE_{DFT}}^\mathrm{PP/XC}}$ for a$_0$\,(\AA) is 0.060(0.051)\%, for E$_\mathrm{coh}$\,(eV/atom) it is 0.375(0.515)\%, for \textbf{$\gamma$}\,(J/m$^2$) is 0.934(1.488)\%, for elastic constants C$_\mathrm{ij}$\,(GPa) and bulk modulus \textbf{B}\,(GPa) it is about 0.6(0.4)\%, and the average for all properties \textbf{$\overline{\sum}$} is 0.522(0.711)\%. \comm{0.454(0.433)\%}
	
	When PP is generated for SCAN(MGGA) and XC functional is PBE or PP is PBE and XC functional is SCAN, then the ${\mathrm{MAPE_{DFT}}^\mathrm{PP/XC}}$ for a$_0$\,(\AA) is 0.831(0.932)\%, for E$_\mathrm{coh}$\,(eV/atom) it is 2.076(1.719)\%, for \textbf{$\gamma$}\,(J/m$^2$) is 4.459(5.666)\%, for elastic constants C$_\mathrm{ij}$\,(GPa) and bulk modulus \textbf{B}\,(GPa) it is about 4(4)\%, and the average for all properties \textbf{$\overline{\sum}$} is 3.401(4.281)\%. \comm{3.225(3.977)\%}
	
	In summary, lattice constants are much less, about 3-5 times, affected by PP and XC functional inconsistency error than other mechanical properties. On average, SCAN/PBE inconsistency error is about 8 times PBE/PBEsol inconsistency and 4 times LDA/GGA inconsistency.
	
	{The average difference of the computed results for different all-electron DFT codes measured by the $\Delta$ gauge (Eq.\ref{eq:deltapp}) is between 0.3 and 1\,meV/atom, while the difference between the results of plane-wave codes using pseudopotentials, coherent with the PBE XC functional, and the all-electron approach is between 0.3 (latest pseudizations) and 13.5\,meV/atom\cite{Lejaeghere2016}. 
In our case, the inconsistency error measured similarly is about 2\,meV/atom for PW LDA/GGA calculations and about 0.5\,meV/atom for GGA$^1$/GGA$^2$ (1$\neq$2), and about 6\,meV/atom for AO MGGA/GGA calculations, see Tab.\ref{tab:meanpp}. As you can see, these are significant, non-negligible quantities.
}

A secondary {and less important} goal of this work was also to evaluate the performance of the XC functionals used here for DFT calculations in reproducing experimental values of structural and mechanical properties of crystals (lattice constant a$_0$\,(\AA), cohesive energy E$_\mathrm{coh}$\,(eV/atom), and  bulk modulus \textbf{B}\,(GPa)). 
Mean absolute percentage error ${\mathrm{MAPE_{EXP}}^\mathrm{PP/XC}}$\,(Eq.\ref{eq:mapee}) for results for compatible PPs and XC functionals are summarized in Tab.\ref{tab:meanexp}. From the point of view of the average for all properties \textbf{$\overline{\sum}$}, PBEsol significantly outperforms PBE, which slightly outperforms LDA (\textbf{$\overline{\sum}$$_{\mathrm{PBEsol}}$}=4.503(4.598), \textbf{$\overline{\sum}$$_{\mathrm{PBE}}$}=6.762(6.416), \textbf{$\overline{\sum}$$_{\mathrm{LDA}}$}=7.305(6.564)). Other researchers have obtained similar results\cite{Sun2011} for LDA and GGA XC functionals.
According to the methodology used here, on average for all properties combined SCAN(MGGA) does not outperform PBE (\textbf{$\overline{\sum}$$_{\mathrm{SCAN}}$}=9.892(10.567), \textbf{$\overline{\sum}$$_{\mathrm{PBE^*}}$}=9.209(8.753)). This seemingly surprising finding is the result of SCAN's poor performance in reproducing the cohesive energy; for lattice constant and bulk modulus, SCAN significantly outperforms PBE.
For the full-potential (linearized) augmented plane-wave plus local orbitals method used to solve the Kohn–Sham equations, but without self-consistently for the MGGA functionals, the SCAN XC functional slightly outperforms even PBEsol\cite{Tran2016}. However, in more recent calculations performed with VASP using PP generated for PBE\cite{Parga2023} for bulk transition metals, SCAN was found to be less accurate than AM05(GGA)\cite{Armiento2005}, which is similar to the PBEsol XC functional used here, not only in cohesive energy but also in lattice constant and bulk modulus.

\begin{table}[H]
	\caption{Mean absolute percentage error ${\mathrm{MAPE_{DFT}}^\mathrm{PP/XC}}$\,(Eq.\ref{eq:mapepp}) of lattice constant a$_0$\,(\AA), cohesive energy E$_\mathrm{coh}$\,(eV/atom), elastic constants C$_\mathrm{ij}$\,(GPa), {surface energy \textbf{$\gamma$}\,(J/m$^2$)} and bulk modulus \textbf{B}\,(GPa) {and mean $\overline{\Delta_i}$ gauge\,(meV/atom)\,(Eq.\ref{eq:deltapp})} from \comm{Tables~1--9 in \href{run:./SI.pdf}{Supplementary material}} Tables~\ref{tab:Na} - \ref{tab:LiF} (ALL indicates $\left(\mathrm{LDA^{ABINIT}, PBE^{ABINIT}, PBEsol^{ABINIT}}\right)$, GGA indicates $\left(\mathrm{PBE^{ABINIT}, PBEsol^{ABINIT}}\right)$ and (M)GGA indicates $\left(\mathrm{PBE^{CP2K}, SCAN^{CP2K}}\right)$, the value in brackets stands for the standard deviation)} \label{tab:meanpp}  
	\centering
	\renewcommand{\arraystretch}{1.2}
		\begin{tabular}{|c|c|c|c|}\hline
			& \multicolumn{3}{c|}{${\mathrm{MAPE_{DFT}}^\mathrm{PP/XC}}$}\\\cline{2-4}
			
			& $\mathrm{ALL/ALL}$ & $\mathrm{GGA/GGA}$ & $\mathrm{(M)GGA/(M)GGA}$\\\hline
			a$_0$ & 0.204\,(0.183) & 0.060\,(0.051) & 0.831\,(0.932)\\
			-E$_\mathrm{coh}$ & 0.820\,(0.645) & 0.375\,(0.515) & 2.076\,(1.719)\\
			\textbf{B} & 0.967\,(0.697) & 0.513\,(0.416) & 3.709\,(3.950)\\
			C$_{11}$ & 1.114\,(0.814) & 0.567\,(0.424) & 3.803\,(4.340)\\
			C$_{12}$ & 0.956\,(0.713) & 0.628\,(0.453) & 3.914\,(3.528)\\
			C$_{44}$ & 1.201\,(0.949) & 0.580\,(0.275) & 5.016\,(5.734)\\
			\textbf{$\gamma$} & 2.031\,(2.730) & 0.934\,(1.488) & 4.459\,(5.666)\\\cline{1-4}
			\textbf{$\overline{\sum}$} & 1.042\,(1.322) & 0.522\,(0.711) &	3.401\,(4.281)\\\cline{1-4}
			&\multicolumn{3}{c|}{${\mathrm{\overline{\Delta}_\mathrm{i}(a,b)}}$}\\\cline{2-4}
			& $\mathrm{ALL/ALL}$ & $\mathrm{GGA/GGA}$ & $\mathrm{(M)GGA/(M)GGA}$\\\cline{1-4}
			\textbf{$\overline{\Delta_i}$} & 2.071\,(2.592) & 0.479\,(0.599) & 5.848\,(7.910)\\\cline{1-4}
		\end{tabular}
	\end{table} 
	
	\begin{table}[H]
		\caption{Mean absolute percentage error ${\mathrm{MAPE_{EXP}}^\mathrm{PP/XC}}$\,(Eq.\ref{eq:mapee}) of lattice constant a$_0$\,(\AA), cohesive energy E$_\mathrm{coh}$\,(eV/atom) and bulk modulus \textbf{B}\,(GPa) from \comm{Tables~1--9 in \href{run:./SI.pdf}{Supplementary material}} Tables~\ref{tab:Na} - \ref{tab:LiF} (only results for compatible PP and XC functional are included (those on the diagonal), PBE$^*$ stands for PBE$^\mathrm{CP2K}$, the value in brackets stands for the standard deviation)} \label{tab:meanexp} 
		\centering
		\setlength{\tabcolsep}{4.85pt} 
		\renewcommand{\arraystretch}{1.2}
			\begin{tabular}{|c|c|c|c|c|c|}\hline
				& \multicolumn{5}{c|}{${\mathrm{MAPE_{EXP}}^\mathrm{PP/XC}}$}\\\cline{2-6}
				
				& {$\mathrm{LDA/LDA}$} & {$\mathrm{PBE/PBE}$} & {$\mathrm{PBEsol/PBEsol}$} & {$\mathrm{PBE^*/PBE^*}$} & {$\mathrm{SCAN/SCAN}$}\\\hline
				a$_0$ & 1.463\,(0.972) & 1.295\,(0.825) & 0.360\,(0.245) & 1.611\,(0.784) & 0.829\,(0.511)\\
				-E$_\mathrm{coh}$  & 9.568\,(6.522) & 5.711\,(3.366) & 6.184\,(3.829) & 9.442\,(5.982) & 17.663\,(8.228)\\
				\textbf{B} & 10.885\,(5.805) & 13.279\,(6.165) & 6.964\,(4.761) & 16.575\,(9.023) & 11.183\,(11.085)\\\cline{1-6}
				\textbf{$\overline{\sum}$} & 7.305\,(6.564) & 6.762\,(6.416) & 4.503\,(4.598) & 9.209\,(8.753) & 9.892\,(10.567)\\\cline{1-6}
			\end{tabular}
		\end{table} 	
	
\comm{\begin{tikzpicture}[scale=0.6]
		\foreach \y [count=\n] in {
			{74,25,39,20,3,3,3,3,3},
			{25,53,31,17,7,7,2,3,2},
			{39,31,37,24,3,3,3,3,3},
			{20,17,24,37,2,2,6,5,5},
			{3,7,3,2,12,1,0,0,0},
			{3,7,3,2,1,36,0,0,0},
			{3,2,3,6,0,0,45,1,1},
			{3,3,3,5,0,0,1,23,1},
			{3,2,3,5,0,0,1,1,78},
		} {
			\ifnum\n<10
			\node[minimum size=6mm] at (\n, 0) {\n};
			\fi
			\foreach \x [count=\m] in \y {
				\node[fill=yellow!\x!purple, minimum size=6mm, text=white] at (\m,-\n) {\x};
			}
		}
		
		\foreach \a [count=\i] in {a,b,c,d,e,f,g,h,i} {
			\node[minimum size=6mm] at (0,-\i) {\a};
		}
	\end{tikzpicture}
}

\section{Conclusions}
\label{sec:Conc}

The paper estimates the error in the determination of mechanical and structural properties (lattice constant, cohesive energy, surface energy, elastic constants, and bulk modulus) of crystals calculated by DFT due to the inconsistency of the pseudopotentials and XC functionals used (LDA, PBE, PBEsol, and SCAN). Additionally, the performance of these XC functionals in reproducing experimental data for the aforementioned properties of the tested crystals was estimated. The inconsistency errors are estimated by the mean absolute percentage error ${\mathrm{MAPE_{DFT}}^\mathrm{PP/XC}}$\,(Eq.\ref{eq:mapepp}) and summarized in Tab.\ref{tab:meanpp}, whereas the performance of XC functionals is evaluated using the ${\mathrm{MAPE_{EXP}}^\mathrm{PP/XC}}$\,(Eq.\ref{eq:mapee}), see Tab.\ref{tab:meanexp}.

{\begin{itemize}
		\item The lattice constants are much less affected by the PP and XC functional inconsistency error, about 3-15 times less than other mechanical properties;
		{\item The surface energy is particularly sensitive to the PP and XC functional inconsistency;}
		\item On average, the SCAN/PBE inconsistency error (${\mathrm{MAPE_{DFT}}^\mathrm{PP/XC}}$=3.401(4.281)) \comm{3.225(3.977)} is about 7 \comm{8} times higher than the PBE/PBEsol inconsistency error (${\mathrm{MAPE_{DFT}}^\mathrm{PP/XC}}$=0.522(0.711)) \comm{0.454(0.433)} and 3 \comm{4} times higher than the LDA/GGA inconsistency error (${\mathrm{MAPE_{DFT}}^\mathrm{PP/XC}}$= 1.042(1.322)); \comm{0.877(0.779)}
		{\item The average inconsistency error measured by the $\Delta$ gauge is about 2\,meV/atom for PW LDA/GGA calculations and about 0.5\,meV/atom for GGA$^1$/GGA$^2$ (1$\neq$2), and about 6\,meV/atom for AO MGGA/GGA calculations, it is significant, non-negligible};
		\item On average, when reference data comes from an experiment, PBEsol(GGA) XC functional significantly outperforms PBE(GGA), which slightly outperforms LDA{} (\textbf{$\overline{\sum}$$_{\mathrm{PBEsol}}$}=4.503(4.598), \textbf{$\overline{\sum}$$_{\mathrm{PBE}}$}=6.762(6.416), \textbf{$\overline{\sum}$$_{\mathrm{LDA}}$}=7.305(6.564));
		\item On average, when reference data comes from an experiment, SCAN(MGGA) does not outperform PBE(GGA) (\textbf{$\overline{\sum}$$_{\mathrm{SCAN}}$}=9.892(10.567), \textbf{$\overline{\sum}$$_{\mathrm{PBE^*}}$}=9.209(8.753)).
\end{itemize} }

When performing DFT calculations with a given XC functional, it is important to pay attention to the compatibility of the functional with the pseudopotential used. Lack of such compatibility/consistency can result in an average error of about 1\% for LDA/GGA inconsistencies, about 0.5\% for GGA$^1$/GGA$^2$ (1$\neq$2), and about 3\% for MGGA/GGA. All XC functional quality tests/rankings should also include possible inconsistency errors.

\begin{acknowledgement}

\comm{Please use ``The authors thank \ldots'' rather than ``The
authors would like to thank \ldots''.

The author thanks Mats Dahlgren for version one of \textsf{achemso},
and Donald Arseneau for the code taken from \textsf{cite} to move
citations after punctuation. Many users have provided feedback on the
class, which is reflected in all of the different demonstrations
shown in this document. 
}

The author thanks the computing cluster GRAFEN at Biocentrum Ochota, the Interdisciplinary Centre for Mathematical and Computational Modelling of Warsaw University (ICM UW) and Pozna\'n Supercomputing and Networking Center (PSNC) for making its computing resources available.
	
This work was supported by the National Science Centre (NCN -- Poland)  Research Project: No. 2021/43/B/ST8/03207.	
\end{acknowledgement}
		
\bibliography{Refs}

\newpage
		
{\begin{suppinfo}
\label{sec:SI}		

\begin{table}[!ht]
	\caption{Na: lattice constant a$_0$\,(\AA), cohesive energy E$_\mathrm{coh}$\,(eV/atom), elastic constants C$_\mathrm{ij}$\,(GPa), surface energy \textbf{$\gamma$}\,(J/m$^2$), bulk modulus \textbf{B}\,(GPa), absolute percentage error APE$_\mathrm{DFT}^\mathrm{PP/XC}$\,(Eq.\ref{eq:mapepp}) and absolute percentage error APE$_\mathrm{EXP}^\mathrm{PP/XC}$\,(Eq.\ref{eq:mapee}) {and $\Delta_i$ gauge\,(meV/atom)\,(Eq.\ref{eq:deltapp})} (The maximum APEs {and $\Delta_i$s} are marked in gray)} 
	\centering
	\tiny
		\begin{tabular}{|c|c|c|c||c|c|c||c|c|c||}\hline
			a$_0^\mathrm{EXP}$=4.207 & \multicolumn{3}{c||}{DFT} & \multicolumn{3}{c||}{APE$_\mathrm{DFT}^\mathrm{PP/XC}$} & \multicolumn{3}{c||}{APE$_\mathrm{EXP}^\mathrm{PP/XC}$}\\\hline
			\backslashbox{PP~~}{XC}
			& LDA & PBE & PBEsol & LDA & PBE & PBEsol & LDA & PBE & PBEsol\\\hline
			LDA & \textbf{4.068} & 4.220 & 4.191 & 0 & 0.084 & 0.096 & 3.316 & 0.325 & 0.378\\
			PBE & 4.064 & \textbf{4.217} & 4.186 & 0.093 & 0 & 0.020& 3.406 & 0.240 & 0.493\\
			PBEsol & 4.063 & 4.217 & \textbf{4.187} & \cellcolor{lightgray}0.109 & 0.003 & 0 & \cellcolor{lightgray}3.421 & 0.236 & 0.474\\\hline
		\end{tabular}\\[0.5\baselineskip]
		
		\begin{tabular}{|c|c|c||c|c||c|c||}\hline
			a$_0^\mathrm{EXP}$=4.207 & \multicolumn{2}{c||}{DFT} & \multicolumn{2}{c||}{APE$_\mathrm{DFT}^\mathrm{PP/XC}$} & \multicolumn{2}{c||}{APE$_\mathrm{EXP}^\mathrm{PP/XC}$}\\\hline
			\diagbox{PP~~}{XC}
			& PBE & SCAN & PBE & SCAN & PBE & SCAN\\\hline
			PBE & \textbf{4.270} & 4.300 & 0 & 1.801 & 1.502 & \cellcolor{lightgray}2.204\\
			SCAN & 4.192 & \textbf{4.224}& \cellcolor{lightgray}1.840 & 0 & 0.366 & 0.397\\\hline
		\end{tabular}\\[1.5\baselineskip]
		
		\begin{tabular}{|c|c|c|c||c|c|c||c|c|c||}\hline
			-E$_\mathrm{coh}^\mathrm{EXP}$=1.12 & \multicolumn{3}{c||}{DFT} & \multicolumn{3}{c||}{APE$_\mathrm{DFT}^\mathrm{PP/XC}$} & \multicolumn{3}{c||}{APE$_\mathrm{EXP}^\mathrm{PP/XC}$}\\\hline
			\backslashbox{PP~~}{XC}
			& LDA & PBE & PBEsol & LDA & PBE & PBEsol & LDA & PBE & PBEsol\\\hline
			LDA & \textbf{1.175} & 0.997 & 1.082 & 0 & 3.104 & 0.265 & 4.878 & 10.938 & 3.412\\
			PBE & 1.135 & \textbf{1.029} & 1.077 & 3.384 & 0 & 0.156 & 1.329 & 8.085 & 3.817\\
			PBEsol & 1.136 & 0.994 & \textbf{1.079} & 3.253 & \cellcolor{lightgray}3.457 & 0 & 1.466 & \cellcolor{lightgray}11.262 & 3.667\\\hline
		\end{tabular}\\[0.5\baselineskip]
		
		\begin{tabular}{|c|c|c||c|c||c|c||}\hline
			-E$_\mathrm{coh}^\mathrm{EXP}$=1.12 & \multicolumn{2}{c||}{DFT} & \multicolumn{2}{c||}{APE$_\mathrm{DFT}^\mathrm{PP/XC}$} & \multicolumn{2}{c||}{APE$_\mathrm{EXP}^\mathrm{PP/XC}$}\\\hline
			\diagbox{PP~~}{XC}
			& PBE & SCAN & PBE & SCAN & PBE & SCAN\\\hline
			PBE & \textbf{1.244} & 1.301 & 0 & 1.881 & 11.100 & 16.180\\
			SCAN & 1.303 & \textbf{1.326} & \cellcolor{lightgray}4.753 &	0 &	16.381 & \cellcolor{lightgray}18.408\\\hline
		\end{tabular}\\[1.5\baselineskip]	
		
		\comm{\begin{tabular}{|c|c|c|c||c|c|c||}\hline
				C$_\mathrm{11}$ & \multicolumn{3}{c||}{DFT} & \multicolumn{3}{c||}{APE$_\mathrm{DFT}^\mathrm{PP/XC}$}\\\hline
				\backslashbox{PP~~}{XC}
				& LDA & PBE & PBEsol & LDA & PBE & PBEsol\\\hline
				LDA & \textbf{9.462} & 7.995 & 8.155 & 0 & \cellcolor{lightgray}2.545 & 1.145\\
				PBE & 9.411 & \textbf{7.796} & 8.180 & 0.546 & 0 & 0.835\\
				PBEsol & 9.315 & 7.663 & \textbf{8.249} & 1.564 & 1.717 & 0\\\hline
			\end{tabular}\\[0.5\baselineskip]
			
			\begin{tabular}{|c|c|c||c|c||}\hline
				C$_\mathrm{11}$ & \multicolumn{2}{c||}{DFT} & \multicolumn{2}{c||}{APE$_\mathrm{DFT}^\mathrm{PP/XC}$}\\\hline
				\diagbox{PP~~}{XC}
				& PBE & SCAN & PBE & SCAN\\\hline
				PBE & \textbf{7.296} & 7.082 & 0 & \cellcolor{lightgray}4.991\\
				SCAN & 7.623 & \textbf{7.454} & 4.474 & 0\\\hline
			\end{tabular}\\[1.5\baselineskip]
		}
		\begin{tabular}{|c|c|c|c||c|c|c||c|c|c||c|c||}\hline
			C$_\mathrm{11}$ & \multicolumn{3}{c||}{DFT} & \multicolumn{3}{c||}{APE$_\mathrm{DFT}^\mathrm{PP/XC}$} & C$_\mathrm{11}$ & \multicolumn{2}{c||}{DFT}& \multicolumn{2}{c||}{APE$_\mathrm{DFT}^\mathrm{PP/XC}$}\\\hline
			\backslashbox{PP}{XC} & LDA & PBE & PBEsol & LDA & PBE & PBEsol & \backslashbox{PP}{XC} & PBE & SCAN & PBE & SCAN\\\hline
			LDA & \textbf{9.462} & 7.995 & 8.155 & 0 & \cellcolor{lightgray}2.545 & 1.145 & PBE & \textbf{7.296} & 7.082 & 0 & \cellcolor{lightgray}4.991\\
			PBE & 9.411 & \textbf{7.796} & 8.180 & 0.546 & 0 & 0.835 & SCAN & 7.623 & \textbf{7.454} & 4.474 & 0\\\cline{8-12}
			PBEsol & 9.315 & 7.663 & \textbf{8.249} & 1.564 & 1.717 & 0\\\cline{1-7}
		\end{tabular}\\[1.5\baselineskip]
		
		\comm{\begin{tabular}{|c|c|c|c||c|c|c||}\hline
				C$_\mathrm{12}$ & \multicolumn{3}{c||}{DFT} & \multicolumn{3}{c||}{APE$_\mathrm{DFT}^\mathrm{PP/XC}$}\\\hline
				\backslashbox{PP~~}{XC}
				& LDA & PBE & PBEsol & LDA & PBE & PBEsol\\\hline
				LDA & \textbf{8.503} & 7.074 & 7.209 & 0 & 0.269 & \cellcolor{lightgray}0.761\\
				PBE & 8.515 & \textbf{7.093} & 7.254 & 0.148 & 0 & 0.134\\
				PBEsol & 8.492 & 7.073 & \textbf{7.264} & 0.123 & 0.275 & 0\\\hline
			\end{tabular}\\[0.25\baselineskip]
			
			\begin{tabular}{|c|c|c||c|c||}\hline
				C$_\mathrm{12}$ & \multicolumn{2}{c||}{DFT} & \multicolumn{2}{c||}{APE$_\mathrm{DFT}^\mathrm{PP/XC}$}\\\hline
				\diagbox{PP~~}{XC}
				& PBE & SCAN & PBE & SCAN\\\hline
				PBE & \textbf{4.654} & 4.522 & 0 & \cellcolor{lightgray}4.900\\
				SCAN & 4.862 & \textbf{4.755} & 4.477 & 0\\\hline
			\end{tabular}\\[0.5\baselineskip]
		}
		\begin{tabular}{|c|c|c|c||c|c|c||c|c|c||c|c||}\hline
			C$_\mathrm{12}$ & \multicolumn{3}{c||}{DFT} & \multicolumn{3}{c||}{APE$_\mathrm{DFT}^\mathrm{PP/XC}$} & C$_\mathrm{12}$ & \multicolumn{2}{c||}{DFT}& \multicolumn{2}{c||}{APE$_\mathrm{DFT}^\mathrm{PP/XC}$}\\\hline
			\backslashbox{PP}{XC} & LDA & PBE & PBEsol & LDA & PBE & PBEsol & \backslashbox{PP}{XC} & PBE & SCAN & PBE & SCAN\\\hline
			LDA & \textbf{8.503} & 7.074 & 7.209 & 0 & 0.269 & \cellcolor{lightgray}0.761 & PBE & \textbf{4.654} & 4.522 & 0 & \cellcolor{lightgray}4.900\\
			PBE & 8.515 & \textbf{7.093} & 7.254 & 0.148 & 0 & 0.134 & SCAN & 4.862 & \textbf{4.755} & 4.477 & 0\\\cline{8-12}
			PBEsol & 8.492 & 7.073 & \textbf{7.264} & 0.123 & 0.275 & 0\\\cline{1-7}
		\end{tabular}\\[1.5\baselineskip]
		
		\comm{\begin{tabular}{|c|c|c|c||c|c|c||}\hline
				C$_\mathrm{44}$ & \multicolumn{3}{c||}{DFT} & \multicolumn{3}{c||}{APE$_\mathrm{DFT}^\mathrm{PP/XC}$}\\\hline
				\backslashbox{PP~~}{XC}
				& LDA & PBE & PBEsol & LDA & PBE & PBEsol\\\hline
				LDA & \textbf{7.504} & 6.350 & 6.641 & 0 & 0.135 & 0.516\\
				PBE & 7.480 & \textbf{6.341} & 6.657 & 0.323 & 0 & 0.274\\
				PBEsol & 7.461 & 6.326 & \textbf{6.676} & \cellcolor{lightgray}0.568 & 0.232 & 0\\\hline
			\end{tabular}\\[0.25\baselineskip]
			
			\begin{tabular}{|c|c|c||c|c||}\hline
				C$_\mathrm{44}$ & \multicolumn{2}{c||}{DFT} & \multicolumn{2}{c||}{APE$_\mathrm{DFT}^\mathrm{PP/XC}$}\\\hline
				\diagbox{PP~~}{XC}
				& PBE & SCAN & PBE & SCAN\\\hline
				PBE & \textbf{2.363} & 2.292 & 0 & \cellcolor{lightgray}5.909\\
				SCAN & 2.503 & \textbf{2.436} & 5.894 & 0\\\hline
			\end{tabular}\\[0.5\baselineskip]
		}
		\begin{tabular}{|c|c|c|c||c|c|c||c|c|c||c|c||}\hline
			C$_\mathrm{44}$ & \multicolumn{3}{c||}{DFT} & \multicolumn{3}{c||}{APE$_\mathrm{DFT}^\mathrm{PP/XC}$} & C$_\mathrm{44}$ & \multicolumn{2}{c||}{DFT}& \multicolumn{2}{c||}{APE$_\mathrm{DFT}^\mathrm{PP/XC}$}\\\hline
			\backslashbox{PP}{XC} & LDA & PBE & PBEsol & LDA & PBE & PBEsol & \backslashbox{PP}{XC} & PBE & SCAN & PBE & SCAN\\\hline
			LDA & \textbf{7.504} & 6.350 & 6.641 & 0 & 0.135 & 0.516 & PBE & \textbf{2.363} & 2.292 & 0 & \cellcolor{lightgray}5.909\\
			PBE & 7.480 & \textbf{6.341} & 6.657 & 0.323 & 0 & 0.274 & SCAN & 2.503 & \textbf{2.436} & 5.894 & 0\\\cline{8-12}
			PBEsol & 7.461 & 6.326 & \textbf{6.676} & \cellcolor{lightgray}0.568 & 0.232 & 0\\\cline{1-7}
		\end{tabular}\\[1.5\baselineskip]
		{\begin{tabular}{|c|c|c|c||c|c|c||c|c|c||c|c||}\hline
				\textbf{$\gamma$} & \multicolumn{3}{c||}{DFT} & \multicolumn{3}{c||}{APE$_\mathrm{DFT}^\mathrm{PP/XC}$} & \textbf{$\gamma$} & \multicolumn{2}{c||}{DFT}& \multicolumn{2}{c||}{APE$_\mathrm{DFT}^\mathrm{PP/XC}$}\\\hline
				\backslashbox{PP}{XC} & LDA & PBE & PBEsol & LDA & PBE & PBEsol & \backslashbox{PP}{XC} & PBE & SCAN & PBE & SCAN\\\hline
				LDA & \textbf{0.245} & 0.206 & 0.225 & 0 & \cellcolor{lightgray}0.484 & 0.146 & PBE & \textbf{0.235} & 0.248 & 0 & 8.797\\
				PBE & 0.244 & \textbf{0.205} & 0.225 & 0.171 & 0 & 0.113 & SCAN & 0.265& \textbf{0.272} & \cellcolor{lightgray}12.838 & 0\\\cline{8-12}
				PBEsol & 0.245 & 0.205 & \textbf{0.225} & 0.092 & 0.189 & 0\\\cline{1-7}
			\end{tabular}\\[1.5\baselineskip]}
		
		
		\begin{tabular}{|c|c|c|c||c|c|c||c|c|c||}\hline
			\textbf{B}$^\mathrm{EXP}$=7.9 & \multicolumn{3}{c||}{DFT} & \multicolumn{3}{c||}{APE$_\mathrm{DFT}^\mathrm{PP/XC}$} & \multicolumn{3}{c||}{APE$_\mathrm{EXP}^\mathrm{PP/XC}$}\\\hline
			\backslashbox{PP~~}{XC}
			& LDA & PBE & PBEsol & LDA & PBE & PBEsol & LDA & PBE & PBEsol\\\hline
			LDA & \textbf{8.823} & 7.381 & 7.524 & 0 & 0.729 & \cellcolor{lightgray}0.900 & \cellcolor{lightgray}11.680 & 6.574 & 4.761\\
			PBE & 8.814 & \textbf{7.327} & 7.563 & 0.100 & 0 & 0.388 & 11.568 & 7.250 & 4.270\\
			PBEsol & 8.766 & 7.270 & \textbf{7.592} & 0.638 & 0.786 & 0 & 10.967 & 7.979 & 3.897\\\hline
		\end{tabular}\\[0.5\baselineskip]
		
		\begin{tabular}{|c|c|c||c|c||c|c||}\hline
			\textbf{B}$^\mathrm{EXP}$=7.9 & \multicolumn{2}{c||}{DFT} & \multicolumn{2}{c||}{APE$_\mathrm{DFT}^\mathrm{PP/XC}$} & \multicolumn{2}{c||}{APE$_\mathrm{EXP}^\mathrm{PP/XC}$}\\\hline
			\diagbox{PP~~}{XC}
			& PBE & SCAN & PBE & SCAN & PBE & SCAN\\\hline
			PBE & \textbf{5.535} & 5.375 & 0 & \cellcolor{lightgray}4.940 & 29.942 & \cellcolor{lightgray}31.957\\
			SCAN & 5.782 & \textbf{5.655} & 4.476 & 0 & 26.806 &	28.421\\\hline
		\end{tabular}\\[1.5\baselineskip]
		
		\begin{tabular}{|c|c|c|c||c|c|c||}\hline
			\textbf{$\Delta_i$} & \multicolumn{3}{c||}{\textbf{$\Delta_i$}(a,b)} & \multicolumn{3}{c||}{\textbf{$\Delta_i$}(a,b)}\\\hline
			\backslashbox{PP~~}{XC} & LDA & PBE & PBEsol & \backslashbox{PP~~}{XC} & PBE & SCAN\\\hline
			LDA & 0 & 0.155 & 0.173 & PBE & 0 & 2.485\\
			PBE & 0.181 & 0 & 0.037 & SCAN & \cellcolor{lightgray}2.598 & 0 \\\cline{5-7}
			PBEsol & \cellcolor{lightgray}0.213 & 0.013 & 0 \\\cline{1-4}
		\end{tabular}
		
		\label{tab:Na}
	\end{table}
	
	\begin{table}[!ht]
		\caption{Li: lattice constant a$_0$\,(\AA), cohesive energy E$_\mathrm{coh}$\,(eV/atom), elastic constants C$_\mathrm{ij}$\,(GPa), surface energy \textbf{$\gamma$}\,(J/m$^2$), bulk modulus \textbf{B}\,(GPa), absolute percentage error APE$_\mathrm{DFT}^\mathrm{PP/XC}$\,(Eq.\ref{eq:mapepp}) and absolute percentage error APE$_\mathrm{EXP}^\mathrm{PP/XC}$\,(Eq.\ref{eq:mapee}) {and $\Delta_i$ gauge\,(meV/atom)\,(Eq.\ref{eq:deltapp})} (The maximum APEs {and $\Delta_i$s} are marked in gray)} \label{tab:Li}
		\centering
		\tiny
			\begin{tabular}{|c|c|c|c||c|c|c||c|c|c||}\hline
				a$_0^\mathrm{EXP}$=3.451 & \multicolumn{3}{c||}{DFT} & \multicolumn{3}{c||}{APE$_\mathrm{DFT}^\mathrm{PP/XC}$} & \multicolumn{3}{c||}{APE$_\mathrm{EXP}^\mathrm{PP/XC}$}\\\hline
				\backslashbox{PP~~}{XC}
				& LDA & PBE & PBEsol & LDA & PBE & PBEsol & LDA & PBE & PBEsol\\\hline
				LDA & \textbf{3.371} & 3.441 & 3.442 & 0 & 0.098 & 0.031 & \cellcolor{lightgray}2.309 &	0.293 & 0.263\\
				PBE & 3.375 & \textbf{3.444} & 3.445 & \cellcolor{lightgray}0.120 & 0 & 0.072 & 2.192 & 0.195 & 0.161\\
				PBEsol & 3.373 & 3.442 & 3.443 & 0.036 & 0.067 & 0 & 2.275 & 0.262 & 0.232\\\hline
			\end{tabular}\\[0.5\baselineskip]
			
			\begin{tabular}{|c|c|c||c|c||c|c||}\hline
				a$_0^\mathrm{EXP}$=3.451 & \multicolumn{2}{c||}{DFT} & \multicolumn{2}{c||}{APE$_\mathrm{DFT}^\mathrm{PP/XC}$} & \multicolumn{2}{c||}{APE$_\mathrm{EXP}^\mathrm{PP/XC}$}\\\hline
				\diagbox{PP~~}{XC}
				& PBE & SCAN & PBE & SCAN & PBE & SCAN\\\hline
				PBE & \textbf{3.467} & 3.507 & 0 & 0.047 & 0.455 & \cellcolor{lightgray}1.632\\
				SCAN & 3.465 & \textbf{3.506} & \cellcolor{lightgray}0.056 & 0 & 0.398 & 1.584\\\hline
			\end{tabular}\\[1.5\baselineskip]
			
			\begin{tabular}{|c|c|c|c||c|c|c||c|c|c||}\hline
				-E$_\mathrm{coh}^\mathrm{EXP}$=1.67 & \multicolumn{3}{c||}{DFT} & \multicolumn{3}{c||}{APE$_\mathrm{DFT}^\mathrm{PP/XC}$} & \multicolumn{3}{c||}{APE$_\mathrm{EXP}^\mathrm{PP/XC}$}\\\hline
				\backslashbox{PP~~}{XC}
				& LDA & PBE & PBEsol & LDA & PBE & PBEsol & LDA & PBE & PBEsol\\\hline
				LDA & \textbf{1.766} & 1.632 & 1.678 & 0 & 1.290 & 0.100 & \cellcolor{lightgray}5.760 & 2.294 & 0.464\\
				PBE & 1.742 & \textbf{1.611} & 1.673 & \cellcolor{lightgray}1.348 & 0 & 0.194 & 4.335 & 3.539 & 0.169\\
				PBEsol & 1.746 & 1.614 & \textbf{1.676} & 1.124 & 0.200 & 0 & 4.571 & 3.346 & 0.364\\\hline
			\end{tabular}\\[0.5\baselineskip]
			
			\begin{tabular}{|c|c|c||c|c||c|c||}\hline
				-E$_\mathrm{coh}^\mathrm{EXP}$=1.67 & \multicolumn{2}{c||}{DFT} & \multicolumn{2}{c||}{APE$_\mathrm{DFT}^\mathrm{PP/XC}$} & \multicolumn{2}{c||}{APE$_\mathrm{EXP}^\mathrm{PP/XC}$}\\\hline
				\diagbox{PP~~}{XC}
				& PBE & SCAN & PBE & SCAN & PBE & SCAN\\\hline
				PBE & \textbf{1.869} & 1.901 & 0 & \cellcolor{lightgray}2.915 & 11.920 & 13.817\\
				SCAN & 1.870 & \textbf{1.958} & 0.065 & 0 & 11.993 & \cellcolor{lightgray}17.235\\\hline
			\end{tabular}\\[1.5\baselineskip]	
			
			\comm{\begin{tabular}{|c|c|c|c||c|c|c||}\hline
					C$_\mathrm{11}$ & \multicolumn{3}{c||}{DFT} & \multicolumn{3}{c||}{APE$_\mathrm{DFT}^\mathrm{PP/XC}$}\\\hline
					\backslashbox{PP~~}{XC}
					& LDA & PBE & PBEsol & LDA & PBE & PBEsol\\\hline
					LDA & \textbf{22.887} & 21.327 &	20.925 & 0 & \cellcolor{lightgray}1.503 & 0.279\\
					PBE & 22.856 & \textbf{21.011} & 20.744 & 0.135 & 0 & 1.144\\
					PBEsol & 22.938 & 21.269 & \textbf{20.984} &	0.224 &	1.229 &	0\\\hline
				\end{tabular}\\[0.5\baselineskip]
				
				\begin{tabular}{|c|c|c||c|c||}\hline
					C$_\mathrm{11}$ & \multicolumn{2}{c||}{DFT} & \multicolumn{2}{c||}{APE$_\mathrm{DFT}^\mathrm{PP/XC}$}\\\hline
					\diagbox{PP~~}{XC}
					& PBE & SCAN & PBE & SCAN\\\hline
					PBE & \textbf{13.678} & 11.871 & 0 & 0.259\\
					SCAN & 13.753 &	\textbf{11.902} & \cellcolor{lightgray}0.550 & 0\\\hline
				\end{tabular}\\[1.5\baselineskip]
			}
			\begin{tabular}{|c|c|c|c||c|c|c||c|c|c||c|c||}\hline
				C$_\mathrm{11}$ & \multicolumn{3}{c||}{DFT} & \multicolumn{3}{c||}{APE$_\mathrm{DFT}^\mathrm{PP/XC}$} & C$_\mathrm{11}$ & \multicolumn{2}{c||}{DFT}& \multicolumn{2}{c||}{APE$_\mathrm{DFT}^\mathrm{PP/XC}$}\\\hline
				\backslashbox{PP}{XC} & LDA & PBE & PBEsol & LDA & PBE & PBEsol & \backslashbox{PP}{XC} & PBE & SCAN & PBE & SCAN\\\hline
				LDA & \textbf{22.887} & 21.327 &	20.925 & 0 & \cellcolor{lightgray}1.503 & 0.279 & PBE & \textbf{13.678} & 11.871 & 0 & 0.259\\
				PBE & 22.856 & \textbf{21.011} & 20.744 & 0.135 & 0 & 1.144 & SCAN & 13.753 & \textbf{11.902} & \cellcolor{lightgray}0.550 & 0\\\cline{8-12}
				PBEsol & 22.938 & 21.269 & \textbf{20.984} & 0.224 & 1.229 & 0\\\cline{1-7}
			\end{tabular}\\[1.5\baselineskip]
			
			\comm{\begin{tabular}{|c|c|c|c||c|c|c||}\hline
					C$_\mathrm{12}$ & \multicolumn{3}{c||}{DFT} & \multicolumn{3}{c||}{APE$_\mathrm{DFT}^\mathrm{PP/XC}$}\\\hline
					\backslashbox{PP~~}{XC}
					& LDA & PBE & PBEsol & LDA & PBE & PBEsol\\\hline
					LDA & \textbf{12.576} & 11.573 & 11.324 & 0 & \cellcolor{lightgray}1.863 & 0.339\\
					PBE & 12.538 & \textbf{11.362} & 11.179 & 0.301 & 0 & 1.618\\
					PBEsol & 12.581 & 11.569 & \textbf{11.363} & 0.038 & 1.825 &	0\\\hline
				\end{tabular}\\[0.5\baselineskip]
				
				\begin{tabular}{|c|c|c||c|c||}\hline
					C$_\mathrm{12}$ & \multicolumn{2}{c||}{DFT} & \multicolumn{2}{c||}{APE$_\mathrm{DFT}^\mathrm{PP/XC}$}\\\hline
					\diagbox{PP~~}{XC}
					& PBE & SCAN & PBE & SCAN\\\hline
					PBE & \textbf{8.726} & 8.499 & 0 & 0.345\\
					SCAN & 8.766 & \textbf{8.528} & \cellcolor{lightgray}0.455 & 0\\\hline
				\end{tabular}\\[1.5\baselineskip]
			}
			\begin{tabular}{|c|c|c|c||c|c|c||c|c|c||c|c||}\hline
				C$_\mathrm{12}$ & \multicolumn{3}{c||}{DFT} & \multicolumn{3}{c||}{APE$_\mathrm{DFT}^\mathrm{PP/XC}$} & C$_\mathrm{12}$ & \multicolumn{2}{c||}{DFT}& \multicolumn{2}{c||}{APE$_\mathrm{DFT}^\mathrm{PP/XC}$}\\\hline
				\backslashbox{PP}{XC} & LDA & PBE & PBEsol & LDA & PBE & PBEsol & \backslashbox{PP}{XC} & PBE & SCAN & PBE & SCAN\\\hline
				LDA & \textbf{12.576} & 11.573 & 11.324 & 0 & \cellcolor{lightgray}1.863 & 0.339 & PBE & \textbf{8.726} & 8.499 & 0 & 0.345\\
				PBE & 12.538 & \textbf{11.362} & 11.179 & 0.301 & 0 & 1.618 & SCAN & 8.766 & \textbf{8.528} & \cellcolor{lightgray}0.455 & 0\\\cline{8-12}
				PBEsol & 12.581 & 11.569 & \textbf{11.363} & 0.038 & 1.825 & 0\\\cline{1-7}
			\end{tabular}\\[1.5\baselineskip]
			
			\comm{\begin{tabular}{|c|c|c|c||c|c|c||}\hline
					C$_\mathrm{44}$ & \multicolumn{3}{c||}{DFT} & \multicolumn{3}{c||}{APE$_\mathrm{DFT}^\mathrm{PP/XC}$}\\\hline
					\backslashbox{PP~~}{XC}
					& LDA & PBE & PBEsol & LDA & PBE & PBEsol\\\hline
					LDA & \textbf{16.137} & 15.122 & 15.066 & 0 & 0.709 & 0.311\\
					PBE & 16.125 & \textbf{15.016} & 15.002 & 0.078 & 0 & 0.734\\
					PBEsol & 16.155 & \textbf{15.129} & 15.113 & 0.109 & \cellcolor{lightgray}0.755 & 0\\\hline
				\end{tabular}\\[0.5\baselineskip]
				
				\begin{tabular}{|c|c|c||c|c||}\hline
					C$_\mathrm{44}$ & \multicolumn{2}{c||}{DFT} & \multicolumn{2}{c||}{APE$_\mathrm{DFT}^\mathrm{PP/XC}$}\\\hline
					\diagbox{PP~~}{XC}
					& PBE & SCAN & PBE & SCAN\\\hline
					PBE & \textbf{5.475} & 5.979 & 0 & 0.584\\
					SCAN & 5.512 & \textbf{6.014} & \cellcolor{lightgray}0.662 & 0\\\hline
				\end{tabular}\\[0.5\baselineskip]
			}		
			\begin{tabular}{|c|c|c|c||c|c|c||c|c|c||c|c||}\hline
				C$_\mathrm{44}$ & \multicolumn{3}{c||}{DFT} & \multicolumn{3}{c||}{APE$_\mathrm{DFT}^\mathrm{PP/XC}$} & C$_\mathrm{44}$ & \multicolumn{2}{c||}{DFT}& \multicolumn{2}{c||}{APE$_\mathrm{DFT}^\mathrm{PP/XC}$}\\\hline
				\backslashbox{PP}{XC} & LDA & PBE & PBEsol & LDA & PBE & PBEsol & \backslashbox{PP}{XC} & PBE & SCAN & PBE & SCAN\\\hline
				LDA & \textbf{16.137} & 15.122 & 15.066 & 0 & 0.709 & 0.311 & PBE & \textbf{5.475} & 5.979 & 0 & 0.584\\
				PBE & 16.125 & \textbf{15.016} & 15.002 & 0.078 & 0 & 0.734 & SCAN & 5.512 & \textbf{6.014} & \cellcolor{lightgray}0.662 & 0\\\cline{8-12}
				PBEsol & 16.155 & \textbf{15.129} & 15.113 & 0.109 & \cellcolor{lightgray}0.755 & 0\\\cline{1-7}
			\end{tabular}\\[1.5\baselineskip]
			{\begin{tabular}{|c|c|c|c||c|c|c||c|c|c||c|c||}\hline
					\textbf{$\gamma$} & \multicolumn{3}{c||}{DFT} & \multicolumn{3}{c||}{APE$_\mathrm{DFT}^\mathrm{PP/XC}$} & \textbf{$\gamma$} & \multicolumn{2}{c||}{DFT}& \multicolumn{2}{c||}{APE$_\mathrm{DFT}^\mathrm{PP/XC}$}\\\hline
					\backslashbox{PP}{XC} & LDA & PBE & PBEsol & LDA & PBE & PBEsol & \backslashbox{PP}{XC} & PBE & SCAN & PBE & SCAN\\\hline
					LDA & \textbf{0.524} & 0.471 & 0.492 & 0 & 0.172 & 0.109 & PBE & \textbf{0.480} & 0.524 & 0	& \cellcolor{lightgray}0.395\\
					PBE & 0.523 & \textbf{0.472} & 0.471 & 0.245 & 0 & \cellcolor{lightgray}3.990 & SCAN & 0.481 & \textbf{0.522} & 0.151 & 0 \\\cline{8-12}
					PBEsol & 0.522 & 0.466 & \textbf{0.491} & 0.327 & 1.129 & 0\\\cline{1-7}
				\end{tabular}\\[1.5\baselineskip]}
			
			\begin{tabular}{|c|c|c|c||c|c|c||c|c|c||}\hline
				\textbf{B}$^\mathrm{EXP}$=13.1 & \multicolumn{3}{c||}{DFT} & \multicolumn{3}{c||}{APE$_\mathrm{DFT}^\mathrm{PP/XC}$} & \multicolumn{3}{c||}{APE$_\mathrm{EXP}^\mathrm{PP/XC}$}\\\hline
				\backslashbox{PP~~}{XC}
				& LDA & PBE & PBEsol & LDA & PBE & PBEsol & LDA & PBE & PBEsol\\\hline
				LDA & \textbf{16.013} &	14.824 & 14.525 & 0 & \cellcolor{lightgray}1.690 & 0.310 & 22.237 & 13.164 & 10.875\\
				PBE & 15.978 & \textbf{14.578} & 14.367 & 0.222 & 0 & 1.391 & 21.966 & 11.283 & 9.674\\
				PBEsol & 16.033 & 14.802 & 14.570 & 0.126 & 1.539 & 0 & \cellcolor{lightgray}22.391 & 13.000 & 11.220\\\hline
			\end{tabular}\\[0.5\baselineskip]
			
			\begin{tabular}{|c|c|c||c|c||c|c||}\hline
				\textbf{B}$^\mathrm{EXP}$=13.1 & \multicolumn{2}{c||}{DFT} & \multicolumn{2}{c||}{APE$_\mathrm{DFT}^\mathrm{PP/XC}$} & \multicolumn{2}{c||}{APE$_\mathrm{EXP}^\mathrm{PP/XC}$}\\\hline
				\diagbox{PP~~}{XC}
				& PBE & SCAN & PBE & SCAN & PBE & SCAN\\\hline
				PBE & \textbf{10.377} & 9.623 & 0 & 0.310 & 20.787 & \cellcolor{lightgray}26.542\\
				SCAN & 10.428 &	\textbf{9.653} & \cellcolor{lightgray}0.497 & 0 & 20.393 & 26.314\\\hline
			\end{tabular}\\[1.5\baselineskip]
			
			\begin{tabular}{|c|c|c|c||c|c|c||}\hline
				\textbf{$\Delta_i$} & \multicolumn{3}{c||}{\textbf{$\Delta_i$}(a,b)} & \multicolumn{3}{c||}{\textbf{$\Delta_i$}(a,b)}\\\hline
				\backslashbox{PP~~}{XC} & LDA & PBE & PBEsol & \backslashbox{PP~~}{XC} & PBE & SCAN\\\hline
				LDA & 0 & 0.190 & 0.060 & PBE & 0 & 0.063\\
				PBE & \cellcolor{lightgray}0.240 & 0 & 0.138 & SCAN & \cellcolor{lightgray}0.079& 0 \\\cline{5-7}
				PBEsol & 0.072 & 0.130 & 0 \\\cline{1-4}
			\end{tabular}
		\end{table}
		\begin{table}
			\caption{Rh: lattice constant a$_0$\,(\AA), cohesive energy E$_\mathrm{coh}$\,(eV/atom), elastic constants C$_\mathrm{ij}$\,(GPa), surface energy \textbf{$\gamma$}\,(J/m$^2$), bulk modulus \textbf{B}\,(GPa), absolute percentage error APE$_\mathrm{DFT}^\mathrm{PP/XC}$\,(Eq.\ref{eq:mapepp}) and absolute percentage error APE$_\mathrm{EXP}^\mathrm{PP/XC}$\,(Eq.\ref{eq:mapee}) {and $\Delta_i$ gauge\,(meV/atom)\,(Eq.\ref{eq:deltapp})} (The maximum APEs {and $\Delta_i$s} are marked in gray)} \label{tab:Rh}
			\centering
			\tiny
				\begin{tabular}{|c|c|c|c||c|c|c||c|c|c||}\hline
					a$_0^\mathrm{EXP}$=3.794 & \multicolumn{3}{c||}{DFT} & \multicolumn{3}{c||}{APE$_\mathrm{DFT}^\mathrm{PP/XC}$} & \multicolumn{3}{c||}{APE$_\mathrm{EXP}^\mathrm{PP/XC}$}\\\hline
					\backslashbox{PP~~}{XC}
					& LDA & PBE & PBEsol & LDA & PBE & PBEsol & LDA & PBE & PBEsol\\\hline
					LDA & \textbf{3.771} & 3.849 & 3.797 & 0 & 0.045 & 0.033 & 0.594 & 1.438 & 0.088\\
					PBE & 3.773 & \textbf{3.850} &	3.799 & \cellcolor{lightgray}0.048 &   0 & 0.012 & 0.546 & \cellcolor{lightgray}1.483 & 0.133\\
					PBEsol & 3.773 & 3.850 & \textbf{3.799} & 0.034 & 0.012 & 0 & 0.560 & 1.471 & 0.121\\\hline
				\end{tabular}\\[0.5\baselineskip]
				
				\begin{tabular}{|c|c|c||c|c||c|c||}\hline
					a$_0^\mathrm{EXP}$=3.794 & \multicolumn{2}{c||}{DFT} & \multicolumn{2}{c||}{APE$_\mathrm{DFT}^\mathrm{PP/XC}$} & \multicolumn{2}{c||}{APE$_\mathrm{EXP}^\mathrm{PP/XC}$}\\\hline
					\diagbox{PP~~}{XC}
					& PBE & SCAN & PBE & SCAN & PBE & SCAN\\\hline
					PBE & \textbf{3.882} & 3.836 & 0 & \cellcolor{lightgray}0.314 & 2.308 & 1.095\\
					SCAN & 3.892 & \textbf{3.848} &	0.278 &	0 & \cellcolor{lightgray}2.592 & 1.413\\\hline
				\end{tabular}\\[1.5\baselineskip]
				
				\begin{tabular}{|c|c|c|c||c|c|c||c|c|c||}\hline
					-E$_\mathrm{coh}^\mathrm{EXP}$=5.783 & \multicolumn{3}{c||}{DFT} & \multicolumn{3}{c||}{APE$_\mathrm{DFT}^\mathrm{PP/XC}$} & \multicolumn{3}{c||}{APE$_\mathrm{EXP}^\mathrm{PP/XC}$}\\\hline
					\backslashbox{PP~~}{XC}
					& LDA & PBE & PBEsol & LDA & PBE & PBEsol & LDA & PBE & PBEsol\\\hline
					LDA & \textbf{7.001} & 5.528 & 6.387 & 0 & 0.076 & 0.037 & 21.057 &	4.407 & 10.451\\
					PBE & 7.006 & \textbf{5.532} & 6.393 & \cellcolor{lightgray}0.081 & 0 & 0.045 & \cellcolor{lightgray}21.154 &	4.334 & 10.541\\
					PBEsol & 7.003 & 5.530 & \textbf{6.390} & 0.037 & 0.044 & 0 & 21.101 & 4.376 & 10.491\\\hline
				\end{tabular}\\[0.5\baselineskip]
				
				\begin{tabular}{|c|c|c||c|c||c|c||}\hline
					-E$_\mathrm{coh}^\mathrm{EXP}$=5.783 & \multicolumn{2}{c||}{DFT} & \multicolumn{2}{c||}{APE$_\mathrm{DFT}^\mathrm{PP/XC}$} & \multicolumn{2}{c||}{APE$_\mathrm{EXP}^\mathrm{PP/XC}$}\\\hline
					\diagbox{PP~~}{XC}
					& PBE & SCAN & PBE & SCAN & PBE & SCAN\\\hline
					PBE & \textbf{5.992} & 6.776 & 0 & \cellcolor{lightgray}0.901 & 3.617 & \cellcolor{lightgray}17.173\\
					SCAN & 5.940 & \textbf{6.716} &	0.865 & 0 & 2.721 &	16.126\\\hline
				\end{tabular}\\[1.5\baselineskip]	
				
				\comm{\begin{tabular}{|c|c|c|c||c|c|c||}\hline
						C$_\mathrm{11}$ & \multicolumn{3}{c||}{DFT} & \multicolumn{3}{c||}{APE$_\mathrm{DFT}^\mathrm{PP/XC}$}\\\hline
						\backslashbox{PP~~}{XC}
						& LDA & PBE & PBEsol & LDA & PBE & PBEsol\\\hline
						LDA & \textbf{462.245} & 371.558 & 430.819 & 0 & \cellcolor{lightgray}0.670 & 0.190\\
						PBE & 464.420 & \textbf{374.064} & 428.217 &	0.470 & 0 & 0.415\\
						PBEsol & 463.217 & 372.117 & \textbf{430.002} & 0.210 & 0.520 & 0\\\hline
					\end{tabular}\\[0.5\baselineskip]
					
					\begin{tabular}{|c|c|c||c|c||}\hline
						C$_\mathrm{11}$ & \multicolumn{2}{c||}{DFT} & \multicolumn{2}{c||}{APE$_\mathrm{DFT}^\mathrm{PP/XC}$}\\\hline
						\diagbox{PP~~}{XC}
						& PBE & SCAN & PBE & SCAN\\\hline
						PBE & \textbf{369.679} &	417.120 & 0 & 1.335\\
						SCAN & 363.915 & \textbf{422.762} & \cellcolor{lightgray}1.559 & 0\\\hline
					\end{tabular}\\[1.5\baselineskip]
				}		
				\begin{tabular}{|c|c|c|c||c|c|c||c|c|c||c|c||}\hline
					C$_\mathrm{11}$ & \multicolumn{3}{c||}{DFT} & \multicolumn{3}{c||}{APE$_\mathrm{DFT}^\mathrm{PP/XC}$} & C$_\mathrm{11}$ & \multicolumn{2}{c||}{DFT}& \multicolumn{2}{c||}{APE$_\mathrm{DFT}^\mathrm{PP/XC}$}\\\hline
					\backslashbox{PP}{XC} & LDA & PBE & PBEsol & LDA & PBE & PBEsol & \backslashbox{PP}{XC} & PBE & SCAN & PBE & SCAN\\\hline
					LDA & \textbf{462.245} & 371.558 & 430.819 & 0 & \cellcolor{lightgray}0.670 & 0.190 & PBE & \textbf{369.679} &	417.120 & 0 & 1.335\\
					PBE & 464.420 & \textbf{374.064} & 428.217 & 0.470 & 0 & 0.415 & SCAN & 363.915 & \textbf{422.762} & \cellcolor{lightgray}1.559 & 0\\\cline{8-12}
					PBEsol & 463.217 & 372.117 & \textbf{430.002} & 0.210 & 0.520 & 0\\\cline{1-7}
				\end{tabular}\\[1.5\baselineskip]
				
				\comm{\begin{tabular}{|c|c|c|c||c|c|c||}\hline
						C$_\mathrm{12}$ & \multicolumn{3}{c||}{DFT} & \multicolumn{3}{c||}{APE$_\mathrm{DFT}^\mathrm{PP/XC}$}\\\hline
						\backslashbox{PP~~}{XC}
						& LDA & PBE & PBEsol & LDA & PBE & PBEsol\\\hline
						LDA & \textbf{215.826} & 171.232 & 199.903 & 0 & \cellcolor{lightgray}0.598 & 0.242\\
						PBE & 216.451 & \textbf{172.262} & 198.795 & 0.290 &	0 & 0.314\\
						PBEsol & 216.102 & 171.561 & \textbf{199.421} & 0.128 & 0.407 & 0\\\hline
					\end{tabular}\\[0.5\baselineskip]
					
					\begin{tabular}{|c|c|c||c|c||}\hline
						C$_\mathrm{12}$ & \multicolumn{2}{c||}{DFT} & \multicolumn{2}{c||}{APE$_\mathrm{DFT}^\mathrm{PP/XC}$}\\\hline
						\diagbox{PP~~}{XC}
						& PBE & SCAN & PBE & SCAN\\\hline
						PBE & \textbf{187.499} &	207.570 & 0 & 1.017\\
						SCAN & 184.272 & \textbf{209.702} & \cellcolor{lightgray}1.722 & 0\\\hline
					\end{tabular}\\[1.5\baselineskip]
				}		
				\begin{tabular}{|c|c|c|c||c|c|c||c|c|c||c|c||}\hline
					C$_\mathrm{12}$ & \multicolumn{3}{c||}{DFT} & \multicolumn{3}{c||}{APE$_\mathrm{DFT}^\mathrm{PP/XC}$} & C$_\mathrm{12}$ & \multicolumn{2}{c||}{DFT}& \multicolumn{2}{c||}{APE$_\mathrm{DFT}^\mathrm{PP/XC}$}\\\hline
					\backslashbox{PP}{XC} & LDA & PBE & PBEsol & LDA & PBE & PBEsol & \backslashbox{PP}{XC} & PBE & SCAN & PBE & SCAN\\\hline
					LDA & \textbf{215.826} & 171.232 & 199.903 & 0 & \cellcolor{lightgray}0.598 & 0.242 & PBE & \textbf{187.499} &	207.570 & 0 & 1.017\\
					PBE & 216.451 & \textbf{172.262} & 198.795 & 0.290 & 0 & 0.314 & SCAN & 184.272 & \textbf{209.702} & \cellcolor{lightgray}1.722 & 0\\\cline{8-12}
					PBEsol & 216.102 & 171.561 & \textbf{199.421} & 0.128 & 0.407 & 0\\\cline{1-7}
				\end{tabular}\\[1.5\baselineskip]
				
				\comm{\begin{tabular}{|c|c|c|c||c|c|c||}\hline
						C$_\mathrm{44}$ & \multicolumn{3}{c||}{DFT} & \multicolumn{3}{c||}{APE$_\mathrm{DFT}^\mathrm{PP/XC}$}\\\hline
						\backslashbox{PP~~}{XC}
						& LDA & PBE & PBEsol & LDA & PBE & PBEsol\\\hline
						LDA & \textbf{200.648} & 165.241 & 191.313 & 0 & \cellcolor{lightgray}0.607 & 0.263\\
						PBE & 201.288 & \textbf{166.250} & 190.200 & 0.319 & 0 & 0.320\\
						PBEsol & 200.917 & 165.541 & \textbf{190.811} & 0.134 & 0.427 & 0\\\hline
					\end{tabular}\\[0.5\baselineskip]
					
					\begin{tabular}{|c|c|c||c|c||}\hline
						C$_\mathrm{44}$ & \multicolumn{2}{c||}{DFT} & \multicolumn{2}{c||}{APE$_\mathrm{DFT}^\mathrm{PP/XC}$}\\\hline
						\diagbox{PP~~}{XC}
						& PBE & SCAN & PBE & SCAN\\\hline
						PBE & \textbf{122.866} & 145.191 & 0 & 1.064\\
						SCAN & 121.322 & \textbf{146.752} & \cellcolor{lightgray}1.256 & 0\\\hline
					\end{tabular}\\[1.5\baselineskip]
				}		
				\begin{tabular}{|c|c|c|c||c|c|c||c|c|c||c|c||}\hline
					C$_\mathrm{44}$ & \multicolumn{3}{c||}{DFT} & \multicolumn{3}{c||}{APE$_\mathrm{DFT}^\mathrm{PP/XC}$} & C$_\mathrm{44}$ & \multicolumn{2}{c||}{DFT}& \multicolumn{2}{c||}{APE$_\mathrm{DFT}^\mathrm{PP/XC}$}\\\hline
					\backslashbox{PP}{XC} & LDA & PBE & PBEsol & LDA & PBE & PBEsol & \backslashbox{PP}{XC} & PBE & SCAN & PBE & SCAN\\\hline
					LDA & \textbf{200.648} & 165.241 & 191.313 & 0 & \cellcolor{lightgray}0.607 & 0.263 & PBE & \textbf{122.866} & 145.191 & 0 & 1.064\\
					PBE & 201.288 & \textbf{166.250} & 190.200 & 0.319 & 0 & 0.320 & SCAN & 121.322 & \textbf{146.752} & \cellcolor{lightgray}1.256 & 0\\\cline{8-12}
					PBEsol & 200.917 & 165.541 & \textbf{190.811} & 0.134 & 0.427 & 0\\\cline{1-7}
				\end{tabular}\\[1.5\baselineskip]
				{\begin{tabular}{|c|c|c|c||c|c|c||c|c|c||c|c||}\hline
						\textbf{$\gamma$} & \multicolumn{3}{c||}{DFT} & \multicolumn{3}{c||}{APE$_\mathrm{DFT}^\mathrm{PP/XC}$} & \textbf{$\gamma$} & \multicolumn{2}{c||}{DFT}& \multicolumn{2}{c||}{APE$_\mathrm{DFT}^\mathrm{PP/XC}$}\\\hline
						\backslashbox{PP}{XC} & LDA & PBE & PBEsol & LDA & PBE & PBEsol & \backslashbox{PP}{XC} & PBE & SCAN & PBE & SCAN\\\hline
						LDA & \textbf{3.274} & 2.593 & 3.052 & 0 & 5.100 & 10.907 & PBE & \textbf{2.431} & 2.771 & 0 & \cellcolor{lightgray}0.530\\
						PBE & 2.847 & \textbf{2.467} & 2.642 & 13.047 & 0 & 4.023 & SCAN & 2.423 & \textbf{2.786} & 0.347 & 0\\\cline{8-12}
						PBEsol & 2.846 & 2.204 & \textbf{2.752} & \cellcolor{lightgray}13.072 & 10.645 & 0\\\cline{1-7}
					\end{tabular}\\[1.5\baselineskip]}
				
				\begin{tabular}{|c|c|c|c||c|c|c||c|c|c||}\hline
					\textbf{B}$^\mathrm{EXP}$=272.1 & \multicolumn{3}{c||}{DFT} & \multicolumn{3}{c||}{APE$_\mathrm{DFT}^\mathrm{PP/XC}$} & \multicolumn{3}{c||}{APE$_\mathrm{EXP}^\mathrm{PP/XC}$}\\\hline
					\backslashbox{PP~~}{XC}
					& LDA & PBE & PBEsol & LDA & PBE & PBEsol & LDA & PBE & PBEsol\\\hline
					LDA & \textbf{297.966} & 238.007 & 276.875 & 0 & \cellcolor{lightgray}0.635 & 0.215 & 9.506 & \cellcolor{lightgray}12.529 & 1.755\\
					PBE & 299.107 & \textbf{239.530} & 275.269 & 0.383 & 0 & 0.366 & 9.926 & 11.970 & 1.165\\
					PBEsol & 298.474 & 238.413 & \textbf{276.281} & 0.170 & 0.466 & 0 & 9.693 & 12.380 & 1.537\\\hline
				\end{tabular}\\[0.5\baselineskip]
				
				\begin{tabular}{|c|c|c||c|c||c|c||}\hline
					\textbf{B}$^\mathrm{EXP}$=272.1 & \multicolumn{2}{c||}{DFT} & \multicolumn{2}{c||}{APE$_\mathrm{DFT}^\mathrm{PP/XC}$} & \multicolumn{2}{c||}{APE$_\mathrm{EXP}^\mathrm{PP/XC}$}\\\hline
					\diagbox{PP~~}{XC}
					& PBE & SCAN & PBE & SCAN & PBE & SCAN\\\hline
					PBE & \textbf{248.226} & 277.420 & 0 & 1.176 & 8.774 & 1.955\\
					SCAN & 244.153 & \textbf{280.722} & \cellcolor{lightgray}1.641 & 0 & \cellcolor{lightgray}10.271 & 3.169\\\hline
				\end{tabular}\\[1.5\baselineskip]
				
				\begin{tabular}{|c|c|c|c||c|c|c||}\hline
					\textbf{$\Delta_i$} & \multicolumn{3}{c||}{\textbf{$\Delta_i$}(a,b)} & \multicolumn{3}{c||}{\textbf{$\Delta_i$}(a,b)}\\\hline
					\backslashbox{PP~~}{XC} & LDA & PBE & PBEsol & \backslashbox{PP~~}{XC} & PBE & SCAN\\\hline
					LDA & 0 & 1.026 & 0.807 & PBE & 0 & \cellcolor{lightgray}8.201\\
					PBE & \cellcolor{lightgray}1.273 & 0 & 0.304 & SCAN & 6.529	& 0 \\\cline{5-7}
					PBEsol & 0.907 & 0.290 & 0 \\\cline{1-4}
				\end{tabular}
			\end{table}
			\begin{table}
				\caption{C: lattice constant a$_0$\,(\AA), cohesive energy E$_\mathrm{coh}$\,(eV/atom), elastic constants C$_\mathrm{ij}$\,(GPa), surface energy \textbf{$\gamma$}\,(J/m$^2$), bulk modulus \textbf{B}\,(GPa), absolute percentage error APE$_\mathrm{DFT}^\mathrm{PP/XC}$\,(Eq.\ref{eq:mapepp}) and absolute percentage error APE$_\mathrm{EXP}^\mathrm{PP/XC}$\,(Eq.\ref{eq:mapee}) {and $\Delta_i$ gauge\,(meV/atom)\,(Eq.\ref{eq:deltapp})} (The maximum APEs {and $\Delta_i$s} are marked in gray)} \label{tab:C}
				\centering
				\tiny
					\begin{tabular}{|c|c|c|c||c|c|c||c|c|c||}\hline
						a$_0^\mathrm{EXP}$=3.555 & \multicolumn{3}{c||}{DFT} & \multicolumn{3}{c||}{APE$_\mathrm{DFT}^\mathrm{PP/XC}$} & \multicolumn{3}{c||}{APE$_\mathrm{EXP}^\mathrm{PP/XC}$}\\\hline
						\backslashbox{PP~~}{XC}
						& LDA & PBE & PBEsol & LDA & PBE & PBEsol & LDA & PBE & PBEsol\\\hline
						LDA & \textbf{3.538} & 3.556 & 3.541 & 0 & 0.452 & 0.497 & 0.486 & 0.041 & 0.396\\
						PBE & 3.554 & \textbf{3.573} &	3.557 &	0.459 & 0 & 0.045 & 0.029 & 0.496 & 0.057\\
						PBEsol & 3.555 & 3.574 & \textbf{3.559} & \cellcolor{lightgray}0.502 & 0.046 & 0 & 0.014 & \cellcolor{lightgray}0.542 & 0.102\\\hline
					\end{tabular}\\[0.5\baselineskip]
					
					\begin{tabular}{|c|c|c||c|c||c|c||}\hline
						a$_0^\mathrm{EXP}$=3.555 & \multicolumn{2}{c||}{DFT} & \multicolumn{2}{c||}{APE$_\mathrm{DFT}^\mathrm{PP/XC}$} & \multicolumn{2}{c||}{APE$_\mathrm{EXP}^\mathrm{PP/XC}$}\\\hline
						\diagbox{PP~~}{XC}
						& PBE & SCAN & PBE & SCAN & PBE & SCAN\\\hline
						PBE & \textbf{3.577} & 3.567 & 0 & 0.232 & \cellcolor{lightgray}0.620 & 0.349\\
						SCAN & 3.567 & \textbf{3.559} &	\cellcolor{lightgray}0.233 & 0 & 0.385 & 0.117\\\hline
					\end{tabular}\\[1.5\baselineskip]
					
					\begin{tabular}{|c|c|c|c||c|c|c||c|c|c||}\hline
						-E$_\mathrm{coh}^\mathrm{EXP}$=7.550 & \multicolumn{3}{c||}{DFT} & \multicolumn{3}{c||}{APE$_\mathrm{DFT}^\mathrm{PP/XC}$} & \multicolumn{3}{c||}{APE$_\mathrm{EXP}^\mathrm{PP/XC}$}\\\hline
						\backslashbox{PP~~}{XC}
						& LDA & PBE & PBEsol & LDA & PBE & PBEsol & LDA & PBE & PBEsol\\\hline
						LDA & \textbf{8.822} & 7.981 & 8.483 & 0 & \cellcolor{lightgray}2.338 & 2.029 & \cellcolor{lightgray}16.847 & 5.712 & 12.362\\
						PBE & 8.634 & \textbf{7.799} & 8.299 & 2.125 & 0 & 0.189 & 14.364 & 3.297 & 9.919\\
						PBEsol & 8.650 & 7.815 & \textbf{8.315} & 1.947 & 0.205 & 0 & 14.572 & 3.509 & 10.127\\\hline
					\end{tabular}\\[0.5\baselineskip]
					
					\begin{tabular}{|c|c|c||c|c||c|c||}\hline
						-E$_\mathrm{coh}^\mathrm{EXP}$=7.550 & \multicolumn{2}{c||}{DFT} & \multicolumn{2}{c||}{APE$_\mathrm{DFT}^\mathrm{PP/XC}$} & \multicolumn{2}{c||}{APE$_\mathrm{EXP}^\mathrm{PP/XC}$}\\\hline
						\diagbox{PP~~}{XC}
						& PBE & SCAN & PBE & SCAN & PBE & SCAN\\\hline
						PBE & \textbf{8.968} & 9.214 & 0 & 1.513 & 18.779 & \cellcolor{lightgray}22.040\\
						SCAN & 9.137 & \textbf{9.077} & \cellcolor{lightgray}1.885 &	0 & 21.019 & 20.221\\\hline
					\end{tabular}\\[1.5\baselineskip]	
					
					\comm{\begin{tabular}{|c|c|c|c||c|c|c||}\hline
							C$_\mathrm{11}$ & \multicolumn{3}{c||}{DFT} & \multicolumn{3}{c||}{APE$_\mathrm{DFT}^\mathrm{PP/XC}$}\\\hline
							\backslashbox{PP~~}{XC}
							& LDA & PBE & PBEsol & LDA & PBE & PBEsol\\\hline
							LDA & \textbf{1097.913} & 1074.898 & 1092.581 & 0 & 2.292 & \cellcolor{lightgray}2.394\\
							PBE & 1074.317 & \textbf{1050.810} & 1069.023 & 2.149 & 0 & 0.187\\
							PBEsol & 1072.288 & 1048.641 & \textbf{1067.033} & 2.334 & 0.206 & 0\\\hline
						\end{tabular}\\[0.5\baselineskip]
						
						\begin{tabular}{|c|c|c||c|c||}\hline
							C$_\mathrm{11}$ & \multicolumn{2}{c||}{DFT} & \multicolumn{2}{c||}{APE$_\mathrm{DFT}^\mathrm{PP/XC}$}\\\hline
							\diagbox{PP~~}{XC}
							& PBE & SCAN & PBE & SCAN\\\hline
							PBE & \textbf{1042.352} & 1097.633 & 0 & \cellcolor{lightgray}1.558\\
							SCAN & 1058.074 & \textbf{1115.001} & 1.508 & 0\\\hline
						\end{tabular}\\[1.5\baselineskip]
					}
					\begin{tabular}{|c|c|c|c||c|c|c||c|c|c||c|c||}\hline
						C$_\mathrm{11}$ & \multicolumn{3}{c||}{DFT} & \multicolumn{3}{c||}{APE$_\mathrm{DFT}^\mathrm{PP/XC}$} & C$_\mathrm{11}$ & \multicolumn{2}{c||}{DFT}& \multicolumn{2}{c||}{APE$_\mathrm{DFT}^\mathrm{PP/XC}$}\\\hline
						\backslashbox{PP}{XC} & LDA & PBE & PBEsol & LDA & PBE & PBEsol & \backslashbox{PP}{XC} & PBE & SCAN & PBE & SCAN\\\hline
						LDA & \textbf{1097.913} & 1074.898 & 1092.581 & 0 & 2.292 & \cellcolor{lightgray}2.394 & PBE & \textbf{1042.352} & 1097.633 & 0 & \cellcolor{lightgray}1.558\\
						PBE & 1074.317 & \textbf{1050.810} & 1069.023 & 2.149 & 0 & 0.187 & SCAN & 1058.074 & \textbf{1115.001} & 1.508 & 0\\\cline{8-12}
						PBEsol & 1072.288 & 1048.641 & \textbf{1067.033} & 2.334 & 0.206 & 0\\\cline{1-7}
					\end{tabular}\\[1.5\baselineskip]
					
					\comm{\begin{tabular}{|c|c|c|c||c|c|c||}\hline
							C$_\mathrm{12}$ & \multicolumn{3}{c||}{DFT} & \multicolumn{3}{c||}{APE$_\mathrm{DFT}^\mathrm{PP/XC}$}\\\hline
							\backslashbox{PP~~}{XC}
							& LDA & PBE & PBEsol & LDA & PBE & PBEsol\\\hline
							LDA & \textbf{148.245} & 128.162 & 143.905 & 0 & \cellcolor{lightgray}4.063 & 2.695\\
							PBE & 143.409 &	\textbf{123.158} & 139.127 & 3.262 & 0 & 0.715\\
							PBEsol & 144.451 & 124.239 & \textbf{140.129} & 2.560 & 0.877 & 0\\\hline
						\end{tabular}\\[0.5\baselineskip]
						
						\begin{tabular}{|c|c|c||c|c||}\hline
							C$_\mathrm{12}$ & \multicolumn{2}{c||}{DFT} & \multicolumn{2}{c||}{APE$_\mathrm{DFT}^\mathrm{PP/XC}$}\\\hline
							\diagbox{PP~~}{XC}
							& PBE & SCAN & PBE & SCAN\\\hline
							PBE & \textbf{112.415} & 105.377 & 0 & \cellcolor{lightgray}3.611\\
							SCAN & 116.097 & \textbf{109.324} & 3.275 & 0\\\hline
						\end{tabular}\\[1.5\baselineskip]
					}	
					\begin{tabular}{|c|c|c|c||c|c|c||c|c|c||c|c||}\hline
						C$_\mathrm{12}$ & \multicolumn{3}{c||}{DFT} & \multicolumn{3}{c||}{APE$_\mathrm{DFT}^\mathrm{PP/XC}$} & C$_\mathrm{12}$ & \multicolumn{2}{c||}{DFT}& \multicolumn{2}{c||}{APE$_\mathrm{DFT}^\mathrm{PP/XC}$}\\\hline
						\backslashbox{PP}{XC} & LDA & PBE & PBEsol & LDA & PBE & PBEsol & \backslashbox{PP}{XC} & PBE & SCAN & PBE & SCAN\\\hline
						LDA & \textbf{148.245} & 128.162 & 143.905 & 0 & \cellcolor{lightgray}4.063 & 2.695 & PBE & \textbf{112.415} & 105.377 & 0 & \cellcolor{lightgray}3.611\\
						PBE & 143.409 &	\textbf{123.158} & 139.127 & 3.262 & 0 & 0.715 & SCAN & 116.097 & \textbf{109.324} & 3.275 & 0\\\cline{8-12}
						PBEsol & 144.451 & 124.239 & \textbf{140.129} & 2.560 & 0.877 & 0\\\cline{1-7}
					\end{tabular}\\[1.5\baselineskip]
					
					\comm{\begin{tabular}{|c|c|c|c||c|c|c||}\hline
							C$_\mathrm{44}$ & \multicolumn{3}{c||}{DFT} & \multicolumn{3}{c||}{APE$_\mathrm{DFT}^\mathrm{PP/XC}$}\\\hline
							\backslashbox{PP~~}{XC}
							& LDA & PBE & PBEsol & LDA & PBE & PBEsol\\\hline
							LDA & \textbf{598.695} & 576.674 & 588.261 & 0 & \cellcolor{lightgray}3.036 & 1.478\\
							PBE & 584.232 & \textbf{559.683} & 574.661 & 2.416 & 0 & 0.868\\
							PBEsol & 583.737 & 561.992 & \textbf{579.693} &	2.498 & 0.413 & 0\\\hline
						\end{tabular}\\[0.5\baselineskip]
						
						\begin{tabular}{|c|c|c||c|c||}\hline
							C$_\mathrm{44}$ & \multicolumn{2}{c||}{DFT} & \multicolumn{2}{c||}{APE$_\mathrm{DFT}^\mathrm{PP/XC}$}\\\hline
							\diagbox{PP~~}{XC}
							& PBE & SCAN & PBE & SCAN\\\hline
							PBE & \textbf{555.279} & 577.077 & 0 & 1.715\\
							SCAN & 565.234 & \textbf{587.148} & \cellcolor{lightgray}1.793 & 0\\\hline
						\end{tabular}\\[1.5\baselineskip]
					}		
					\begin{tabular}{|c|c|c|c||c|c|c||c|c|c||c|c||}\hline
						C$_\mathrm{44}$ & \multicolumn{3}{c||}{DFT} & \multicolumn{3}{c||}{APE$_\mathrm{DFT}^\mathrm{PP/XC}$} & C$_\mathrm{44}$ & \multicolumn{2}{c||}{DFT}& \multicolumn{2}{c||}{APE$_\mathrm{DFT}^\mathrm{PP/XC}$}\\\hline
						\backslashbox{PP}{XC} & LDA & PBE & PBEsol & LDA & PBE & PBEsol & \backslashbox{PP}{XC} & PBE & SCAN & PBE & SCAN\\\hline
						LDA & \textbf{598.695} & 576.674 & 588.261 & 0 & \cellcolor{lightgray}3.036 & 1.478 & PBE & \textbf{555.279} & 577.077 & 0 & 1.715\\
						PBE & 584.232 & \textbf{559.683} & 574.661 & 2.416 & 0 & 0.868 & SCAN & 565.234 & \textbf{587.148} & \cellcolor{lightgray}1.793 & 0\\\cline{8-12}
						PBEsol & 583.737 & 561.992 & \textbf{579.693} &	2.498 & 0.413 & 0\\\cline{1-7}
					\end{tabular}\\[1.5\baselineskip]
					{\begin{tabular}{|c|c|c|c||c|c|c||c|c|c||c|c||}\hline
							\textbf{$\gamma$} & \multicolumn{3}{c||}{DFT} & \multicolumn{3}{c||}{APE$_\mathrm{DFT}^\mathrm{PP/XC}$} & \textbf{$\gamma$} & \multicolumn{2}{c||}{DFT}& \multicolumn{2}{c||}{APE$_\mathrm{DFT}^\mathrm{PP/XC}$}\\\hline
							\backslashbox{PP}{XC} & LDA & PBE & PBEsol & LDA & PBE & PBEsol & \backslashbox{PP}{XC} & PBE & SCAN & PBE & SCAN\\\hline
							LDA & \textbf{9.590} & 8.935 & 9.397 & 0 & 2.882 & 2.675 & PBE & \textbf{8.895} & 9.853 & 0 & 1.835\\
							PBE & 9.332 & \textbf{8.685} & 9.142 & \cellcolor{lightgray}2.689 & 0 & 0.108 & SCAN & 9.065 & \textbf{9.675} & \cellcolor{lightgray}1.914 & 0\\\cline{8-12}
							PBEsol & 9.342 & 8.695 & \textbf{9.152} & 2.584 & 0.120 & 0\\\cline{1-7}
						\end{tabular}\\[1.5\baselineskip]}
					
					\begin{tabular}{|c|c|c|c||c|c|c||c|c|c||}\hline
						\textbf{B}$^\mathrm{EXP}$=454.7 & \multicolumn{3}{c||}{DFT} & \multicolumn{3}{c||}{APE$_\mathrm{DFT}^\mathrm{PP/XC}$} & \multicolumn{3}{c||}{APE$_\mathrm{EXP}^\mathrm{PP/XC}$}\\\hline
						\backslashbox{PP~~}{XC}
						& LDA & PBE & PBEsol & LDA & PBE & PBEsol & LDA & PBE & PBEsol\\\hline
						LDA & \textbf{464.801} & 443.741 & 460.130 & 0 & \cellcolor{lightgray}2.629 & 2.457 & 2.221 & 2.410 & 1.194\\
						PBE & 453.712 & \textbf{432.375} & 449.092 & 2.386 & 0 & 0.001 & 0.217 & 4.910 & 1.233\\
						PBEsol & 453.730 & 432.373 & \textbf{449.097} &	2.382 & 0.001 & 0 & 0.213 & \cellcolor{lightgray}4.910 & 1.232\\\hline
					\end{tabular}\\[0.5\baselineskip]
					
					\begin{tabular}{|c|c|c||c|c||c|c||}\hline
						\textbf{B}$^\mathrm{EXP}$=454.7& \multicolumn{2}{c||}{DFT} & \multicolumn{2}{c||}{APE$_\mathrm{DFT}^\mathrm{PP/XC}$} & \multicolumn{2}{c||}{APE$_\mathrm{EXP}^\mathrm{PP/XC}$}\\\hline
						\diagbox{PP~~}{XC}
						& PBE & SCAN & PBE & SCAN & PBE & SCAN\\\hline
						PBE & \textbf{422.394} & 436.129 & 0 & \cellcolor{lightgray}1.894 & \cellcolor{lightgray}7.105 & 4.084\\
						SCAN & 430.089 & \textbf{444.550} &	1.822 & 0 & 5.413 & 2.232\\\hline
					\end{tabular}\\[1.5\baselineskip]
					
					\begin{tabular}{|c|c|c|c||c|c|c||}\hline
						\textbf{$\Delta_i$} & \multicolumn{3}{c||}{\textbf{$\Delta_i$}(a,b)} & \multicolumn{3}{c||}{\textbf{$\Delta_i$}(a,b)}\\\hline
						\backslashbox{PP~~}{XC} & LDA & PBE & PBEsol & \backslashbox{PP~~}{XC} & PBE & SCAN\\\hline
						LDA & 0 & 7.316 & 8.247 & PBE & 0 & \cellcolor{lightgray}3.742\\
						PBE & 7.632 & 0 & 0.742 & SCAN & 3.691 & 0 \\\cline{5-7}
						PBEsol & \cellcolor{lightgray}8.359 & 0.742 & 0 \\\cline{1-4}
					\end{tabular}
				\end{table}
				\begin{table}
					\caption{Si: lattice constant a$_0$\,(\AA), cohesive energy E$_\mathrm{coh}$\,(eV/atom), elastic constants C$_\mathrm{ij}$\,(GPa), surface energy \textbf{$\gamma$}\,(J/m$^2$), bulk modulus \textbf{B}\,(GPa), absolute percentage error APE$_\mathrm{DFT}^\mathrm{PP/XC}$\,(Eq.\ref{eq:mapepp}) and absolute percentage error APE$_\mathrm{EXP}^\mathrm{PP/XC}$\,(Eq.\ref{eq:mapee}) {and $\Delta_i$ gauge\,(meV/atom)\,(Eq.\ref{eq:deltapp})} (The maximum APEs {and $\Delta_i$s} are marked in gray)} \label{tab:Si}
					\centering
					\tiny
						\begin{tabular}{|c|c|c|c||c|c|c||c|c|c||}\hline
							a$_0^\mathrm{EXP}$=5.422 & \multicolumn{3}{c||}{DFT} & \multicolumn{3}{c||}{APE$_\mathrm{DFT}^\mathrm{PP/XC}$} & \multicolumn{3}{c||}{APE$_\mathrm{EXP}^\mathrm{PP/XC}$}\\\hline
							\backslashbox{PP~~}{XC}
							& LDA & PBE & PBEsol & LDA & PBE & PBEsol & LDA & PBE & PBEsol\\\hline
							LDA & \textbf{5.394} & 5.417 & 5.393 & 0 & 0.949 & 0.695 & 0.512 & 0.096 & 0.528\\
							PBE & 5.446 & \textbf{5.469} &	5.445 & \cellcolor{lightgray}0.956 & 0 & 0.258 & 0.439 & \cellcolor{lightgray}0.862 & 0.428\\
							PBEsol & 5.432 & 5.455 & \textbf{5.431} & 0.701 & 0.258 & 0 & 0.185 & 0.601 & 0.169\\\hline
						\end{tabular}\\[0.5\baselineskip]
						
						\begin{tabular}{|c|c|c||c|c||c|c||}\hline
							a$_0^\mathrm{EXP}$=5.422 & \multicolumn{2}{c||}{DFT} & \multicolumn{2}{c||}{APE$_\mathrm{DFT}^\mathrm{PP/XC}$} & \multicolumn{2}{c||}{APE$_\mathrm{EXP}^\mathrm{PP/XC}$}\\\hline
							\diagbox{PP~~}{XC}
							& PBE & SCAN & PBE & SCAN & PBE & SCAN\\\hline
							PBE & \textbf{5.465} & 5.435 & 0 & 0.450 & \cellcolor{lightgray}0.795 & 0.244\\
							SCAN & 5.440 & \textbf{5.411} & \cellcolor{lightgray}0.453 & 0 & 0.338 & 0.205\\\hline
						\end{tabular}\\[1.5\baselineskip]
						
						\begin{tabular}{|c|c|c|c||c|c|c||c|c|c||}\hline
							-E$_\mathrm{coh}^\mathrm{EXP}$=4.680 & \multicolumn{3}{c||}{DFT} & \multicolumn{3}{c||}{APE$_\mathrm{DFT}^\mathrm{PP/XC}$} & \multicolumn{3}{c||}{APE$_\mathrm{EXP}^\mathrm{PP/XC}$}\\\hline
							\backslashbox{PP~~}{XC}
							& LDA & PBE & PBEsol & LDA & PBE & PBEsol & LDA & PBE & PBEsol\\\hline
							LDA & \textbf{5.349} & 4.872 & 5.191 & 0 & \cellcolor{lightgray}1.589 & 1.051 & \cellcolor{lightgray}14.284 & 4.107 & 10.909\\
							PBE & 5.264 & \textbf{4.796} & 5.109 & 1.578 & 0 & 0.532 & 12.481 & 2.478 & 9.171\\
							PBEsol & 5.292 & 4.822 & \textbf{5.137} & 1.051 & 0.543 & 0 & 13.083 & 3.035 & 9.756\\\hline
						\end{tabular}\\[0.25\baselineskip]
						
						\begin{tabular}{|c|c|c||c|c||c|c||}\hline
							-E$_\mathrm{coh}^\mathrm{EXP}$=4.680 & \multicolumn{2}{c||}{DFT} & \multicolumn{2}{c||}{APE$_\mathrm{DFT}^\mathrm{PP/XC}$} & \multicolumn{2}{c||}{APE$_\mathrm{EXP}^\mathrm{PP/XC}$}\\\hline
							\diagbox{PP~~}{XC}
							& PBE & SCAN & PBE & SCAN & PBE & SCAN\\\hline
							PBE & \textbf{5.414} & 5.902 & 0 & 1.098 & 15.690 & 26.112\\
							SCAN & 5.540 & \textbf{5.968} & \cellcolor{lightgray}2.321 & 0 & 18.375 & \cellcolor{lightgray}27.512\\\hline
						\end{tabular}\\[1.5\baselineskip]	
						
						\comm{\begin{tabular}{|c|c|c|c||c|c|c||}\hline
								C$_\mathrm{11}$ & \multicolumn{3}{c||}{DFT} & \multicolumn{3}{c||}{APE$_\mathrm{DFT}^\mathrm{PP/XC}$}\\\hline
								\backslashbox{PP~~}{XC}
								& LDA & PBE & PBEsol & LDA & PBE & PBEsol\\\hline
								LDA & \textbf{154.753} & 154.869 & 155.127 & 0 & \cellcolor{lightgray}4.526 & 3.189\\
								PBE & 148.030 & \textbf{148.163} & 148.359 & 4.344 & 0 & 1.313\\
								PBEsol & 149.968 & 150.137 & 150.333 & 3.092 & 1.333 & 0\\\hline
							\end{tabular}\\[0.5\baselineskip]
							
							\begin{tabular}{|c|c|c||c|c||}\hline
								C$_\mathrm{11}$ & \multicolumn{2}{c||}{DFT} & \multicolumn{2}{c||}{APE$_\mathrm{DFT}^\mathrm{PP/XC}$}\\\hline
								\diagbox{PP~~}{XC}
								& PBE & SCAN & PBE & SCAN\\\hline
								PBE & \textbf{154.674} & 166.816 & 0 & 3.236\\
								SCAN & 160.618 & \textbf{172.396} &	\cellcolor{lightgray}3.843 & 0\\\hline
							\end{tabular}\\[1.5\baselineskip]
						}	
						\begin{tabular}{|c|c|c|c||c|c|c||c|c|c||c|c||}\hline
							C$_\mathrm{11}$ & \multicolumn{3}{c||}{DFT} & \multicolumn{3}{c||}{APE$_\mathrm{DFT}^\mathrm{PP/XC}$} & C$_\mathrm{11}$ & \multicolumn{2}{c||}{DFT}& \multicolumn{2}{c||}{APE$_\mathrm{DFT}^\mathrm{PP/XC}$}\\\hline
							\backslashbox{PP}{XC} & LDA & PBE & PBEsol & LDA & PBE & PBEsol & \backslashbox{PP}{XC} & PBE & SCAN & PBE & SCAN\\\hline
							LDA & \textbf{154.753} & 154.869 & 155.127 & 0 & \cellcolor{lightgray}4.526 & 3.189 & PBE & \textbf{154.674} & 166.816 & 0 & 3.236\\
							PBE & 148.030 & \textbf{148.163} & 148.359 & 4.344 & 0 & 1.313 & SCAN & 160.618 & \textbf{172.396} &	\cellcolor{lightgray}3.843 & 0\\\cline{8-12}
							PBEsol & 149.968 & 150.137 & 150.333 & 3.092 & 1.333 & 0 \\\cline{1-7}
						\end{tabular}\\[1.5\baselineskip]
						
						\comm{\begin{tabular}{|c|c|c|c||c|c|c||}\hline
								C$_\mathrm{12}$ & \multicolumn{3}{c||}{DFT} & \multicolumn{3}{c||}{APE$_\mathrm{DFT}^\mathrm{PP/XC}$}\\\hline
								\backslashbox{PP~~}{XC}
								& LDA & PBE & PBEsol & LDA & PBE & PBEsol\\\hline
								LDA & \textbf{66.362} & 60.011 & 65.942 & 0 & \cellcolor{lightgray}3.096 & 2.130\\
								PBE & 64.365 & \textbf{58.209} & 63.930 & 3.009 & 0 & 0.986\\
								PBEsol & 64.964 & 58.786 & \textbf{64.566} & 2.107 &	0.991 & 0\\\hline
							\end{tabular}\\[0.5\baselineskip]
							
							\begin{tabular}{|c|c|c||c|c||}\hline
								C$_\mathrm{12}$ & \multicolumn{2}{c||}{DFT} & \multicolumn{2}{c||}{APE$_\mathrm{DFT}^\mathrm{PP/XC}$}\\\hline
								\diagbox{PP~~}{XC}
								& PBE & SCAN & PBE & SCAN\\\hline
								PBE & \textbf{53.165} & 56.231 & 0 & 3.809\\
								SCAN & 55.462 & \textbf{58.458} & \cellcolor{lightgray}4.321 & 0\\\hline
							\end{tabular}\\[1.5\baselineskip]
						}	
						\begin{tabular}{|c|c|c|c||c|c|c||c|c|c||c|c||}\hline
							C$_\mathrm{12}$ & \multicolumn{3}{c||}{DFT} & \multicolumn{3}{c||}{APE$_\mathrm{DFT}^\mathrm{PP/XC}$} & C$_\mathrm{12}$ & \multicolumn{2}{c||}{DFT}& \multicolumn{2}{c||}{APE$_\mathrm{DFT}^\mathrm{PP/XC}$}\\\hline
							\backslashbox{PP}{XC} & LDA & PBE & PBEsol & LDA & PBE & PBEsol & \backslashbox{PP}{XC} & PBE & SCAN & PBE & SCAN\\\hline
							LDA & \textbf{66.362} & 60.011 & 65.942 & 0 & \cellcolor{lightgray}3.096 & 2.130 & PBE & \textbf{53.165} & 56.231 & 0 & 3.809\\
							PBE & 64.365 & \textbf{58.209} & 63.930 & 3.009 & 0 & 0.986 & SCAN & 55.462 & \textbf{58.458} & \cellcolor{lightgray}4.321 & 0\\\cline{8-12}
							PBEsol & 64.964 & 58.786 & \textbf{64.566} & 2.107 &	0.991 & 0 \\\cline{1-7}
						\end{tabular}\\[1.5\baselineskip]
						
						\comm{\begin{tabular}{|c|c|c|c||c|c|c||}\hline
								C$_\mathrm{44}$ & \multicolumn{3}{c||}{DFT} & \multicolumn{3}{c||}{APE$_\mathrm{DFT}^\mathrm{PP/XC}$}\\\hline
								\backslashbox{PP~~}{XC}
								& LDA & PBE & PBEsol & LDA & PBE & PBEsol\\\hline
								LDA & \textbf{69.344} & 73.656 & 69.883 & 0 & \cellcolor{lightgray}4.977 & 3.630\\
								PBE & 65.931 & \textbf{70.164} & 66.446 & 4.922 & 0 & 1.466\\
								PBEsol & 66.914 & 71.161 & \textbf{67.435} & 3.504 & 1.421 & 0\\\hline
							\end{tabular}\\[0.5\baselineskip]
							
							\begin{tabular}{|c|c|c||c|c||}\hline
								C$_\mathrm{44}$ & \multicolumn{2}{c||}{DFT} & \multicolumn{2}{c||}{APE$_\mathrm{DFT}^\mathrm{PP/XC}$}\\\hline
								\diagbox{PP~~}{XC}
								& PBE & SCAN & PBE & SCAN\\\hline
								PBE & \textbf{77.131} & 84.367 & 0 & 3.182\\
								SCAN & 79.788 & \textbf{87.139} & \cellcolor{lightgray}3.444 & 0\\\hline
							\end{tabular}\\[1.5\baselineskip]
						}
						\begin{tabular}{|c|c|c|c||c|c|c||c|c|c||c|c||}\hline
							C$_\mathrm{44}$ & \multicolumn{3}{c||}{DFT} & \multicolumn{3}{c||}{APE$_\mathrm{DFT}^\mathrm{PP/XC}$} & C$_\mathrm{44}$ & \multicolumn{2}{c||}{DFT}& \multicolumn{2}{c||}{APE$_\mathrm{DFT}^\mathrm{PP/XC}$}\\\hline
							\backslashbox{PP}{XC} & LDA & PBE & PBEsol & LDA & PBE & PBEsol & \backslashbox{PP}{XC} & PBE & SCAN & PBE & SCAN\\\hline
							LDA & \textbf{69.344} & 73.656 & 69.883 & 0 & \cellcolor{lightgray}4.977 & 3.630 & PBE & \textbf{77.131} & 84.367 & 0 & 3.182\\
							PBE & 65.931 & \textbf{70.164} & 66.446 & 4.922 & 0 & 1.466 & SCAN & 79.788 & \textbf{87.139} & \cellcolor{lightgray}3.444 & 0\\\cline{8-12}
							PBEsol & 66.914 & 71.161 & \textbf{67.435} & 3.504 & 1.421 & 0 \\\cline{1-7}
						\end{tabular}\\[1.5\baselineskip]
						{\begin{tabular}{|c|c|c|c||c|c|c||c|c|c||c|c||}\hline
								\textbf{$\gamma$} & \multicolumn{3}{c||}{DFT} & \multicolumn{3}{c||}{APE$_\mathrm{DFT}^\mathrm{PP/XC}$} & \textbf{$\gamma$} & \multicolumn{2}{c||}{DFT}& \multicolumn{2}{c||}{APE$_\mathrm{DFT}^\mathrm{PP/XC}$}\\\hline
								\backslashbox{PP}{XC} & LDA & PBE & PBEsol & LDA & PBE & PBEsol & \backslashbox{PP}{XC} & PBE & SCAN & PBE & SCAN\\\hline
								LDA & \textbf{2.226} & 2.057 & 2.187 & 0 & \cellcolor{lightgray}3.784 & 2.552 & PBE & \textbf{2.239} & 2.488 & 0 & 4.026\\
								PBE & 2.146 & \textbf{1.982} & 2.108 & 3.606 & 0 & 1.160 & SCAN & 2.336 & \textbf{2.592} & \cellcolor{lightgray}4.318 & 0\\\cline{8-12}
								PBEsol & 2.171 & 2.006 & \textbf{2.133} & 2.485 & 1.200 & 0\\\cline{1-7}
							\end{tabular}\\[1.5\baselineskip]}		
						
						\begin{tabular}{|c|c|c|c||c|c|c||c|c|c||}\hline
							\textbf{B}$^\mathrm{EXP}$=101.3 & \multicolumn{3}{c||}{DFT} & \multicolumn{3}{c||}{APE$_\mathrm{DFT}^\mathrm{PP/XC}$} & \multicolumn{3}{c||}{APE$_\mathrm{EXP}^\mathrm{PP/XC}$}\\\hline
							\backslashbox{PP~~}{XC}
							& LDA & PBE & PBEsol & LDA & PBE & PBEsol & LDA & PBE & PBEsol\\\hline
							LDA & \textbf{95.826} & 91.630 & 95.670 & 0 & \cellcolor{lightgray}3.897 & 2.700 & 5.404 & 9.546 & 5.558\\
							PBE & 92.254 & \textbf{88.194} & 92.073 & 3.728 & 0 & 1.162 & 8.930 & \cellcolor{lightgray}12.938 & 9.109\\
							PBEsol & 93.299 & 89.236 & \textbf{93.155} & 2.637 & 1.182 & 0 & 7.899 & 11.909 & 8.040\\\hline
						\end{tabular}\\[0.5\baselineskip]
						
						\begin{tabular}{|c|c|c||c|c||c|c||}\hline
							\textbf{B}$^\mathrm{EXP}$=101.3 & \multicolumn{2}{c||}{DFT} & \multicolumn{2}{c||}{APE$_\mathrm{DFT}^\mathrm{PP/XC}$} & \multicolumn{2}{c||}{APE$_\mathrm{EXP}^\mathrm{PP/XC}$}\\\hline
							\diagbox{PP~~}{XC}
							& PBE & SCAN & PBE & SCAN & PBE & SCAN\\\hline
							PBE & \textbf{87.001} & 93.093 & 0 & 3.468 & \cellcolor{lightgray}14.115 & 8.102\\
							SCAN & 90.514 & \textbf{96.437} & \cellcolor{lightgray}4.038 & 0 & 10.648 & 4.801\\\hline
						\end{tabular}\\[1.5\baselineskip]
						
						\begin{tabular}{|c|c|c|c||c|c|c||}\hline
							\textbf{$\Delta_i$} & \multicolumn{3}{c||}{\textbf{$\Delta_i$}(a,b)} & \multicolumn{3}{c||}{\textbf{$\Delta_i$}(a,b)}\\\hline
							\backslashbox{PP~~}{XC} & LDA & PBE & PBEsol & \backslashbox{PP~~}{XC} & PBE & SCAN\\\hline
							LDA & 0 & 11.348 & 8.514 & PBE & 0 & \cellcolor{lightgray}5.505\\
							PBE & \cellcolor{lightgray}11.607 & 0 & 3.133 & SCAN & 5.302 & 0 \\\cline{5-7}
							PBEsol & 8.536 & 3.084 & 0 \\\cline{1-4}
						\end{tabular}
					\end{table}
					\begin{table}
						\caption{Ge: lattice constant a$_0$\,(\AA), cohesive energy E$_\mathrm{coh}$\,(eV/atom), elastic constants C$_\mathrm{ij}$\,(GPa), surface energy \textbf{$\gamma$}\,(J/m$^2$), bulk modulus \textbf{B}\,(GPa), absolute percentage error APE$_\mathrm{DFT}^\mathrm{PP/XC}$\,(Eq.\ref{eq:mapepp}) and absolute percentage error APE$_\mathrm{EXP}^\mathrm{PP/XC}$\,(Eq.\ref{eq:mapee}) {and $\Delta_i$ gauge\,(meV/atom)\,(Eq.\ref{eq:deltapp})} (The maximum APEs {and $\Delta_i$s} are marked in gray)} \label{tab:Ge}
						\centering
						\tiny
							\begin{tabular}{|c|c|c|c||c|c|c||c|c|c||}\hline
								a$_0^\mathrm{EXP}$=5.644 & \multicolumn{3}{c||}{DFT} & \multicolumn{3}{c||}{APE$_\mathrm{DFT}^\mathrm{PP/XC}$} & \multicolumn{3}{c||}{APE$_\mathrm{EXP}^\mathrm{PP/XC}$}\\\hline
								\backslashbox{PP~~}{XC}
								& LDA & PBE & PBEsol & LDA & PBE & PBEsol & LDA & PBE & PBEsol\\\hline
								LDA & \textbf{5.623} & 5.790 & 5.703 & 0 & 0.360 & 0.428 & 0.372 & \cellcolor{lightgray}2.585 & 1.053\\
								PBE & 5.600 & \textbf{5.769} & 5.681 & 0.407 & 0 & 0.031 & 0.777 & 2.217 & 0.654\\
								PBEsol & 5.598 & 5.767 & \textbf{5.679} & \cellcolor{lightgray}0.437 & 0.036 & 0 & 0.808 & 2.180 & 0.623\\\hline
							\end{tabular}\\[0.5\baselineskip]
							
							\begin{tabular}{|c|c|c||c|c||c|c||}\hline
								a$_0^\mathrm{EXP}$=5.644 & \multicolumn{2}{c||}{DFT} & \multicolumn{2}{c||}{APE$_\mathrm{DFT}^\mathrm{PP/XC}$} & \multicolumn{2}{c||}{APE$_\mathrm{EXP}^\mathrm{PP/XC}$}\\\hline
								\diagbox{PP~~}{XC}
								& PBE & SCAN & PBE & SCAN & PBE & SCAN\\\hline
								PBE & \textbf{5.771} & 5.740 & 0 & \cellcolor{lightgray}2.791 & \cellcolor{lightgray}2.245 & 1.693\\
								SCAN & 5.618 & \textbf{5.584} & 2.653 & 0 & 0.468 & 1.068\\\hline
							\end{tabular}\\[1.5\baselineskip]
							
							\begin{tabular}{|c|c|c|c||c|c|c||c|c|c||}\hline
								-E$_\mathrm{coh}^\mathrm{EXP}$=3.890 & \multicolumn{3}{c||}{DFT} & \multicolumn{3}{c||}{APE$_\mathrm{DFT}^\mathrm{PP/XC}$} & \multicolumn{3}{c||}{APE$_\mathrm{EXP}^\mathrm{PP/XC}$}\\\hline
								\backslashbox{PP~~}{XC}
								& LDA & PBE & PBEsol & LDA & PBE & PBEsol & LDA & PBE & PBEsol\\\hline
								LDA & \textbf{4.360} & 3.661 & 4.063 & 0 & 0.320 & 0.564 & 12.083 & 5.876 & 4.449\\
								PBE & 4.387 & \textbf{3.673} & 4.081 & 0.612 & 0 & 0.114 & 12.769 & 5.574 & 4.921\\
								PBEsol & 4.391 & 3.678 & \textbf{4.086}	& \cellcolor{lightgray}0.711 & 0.134 & 0 & \cellcolor{lightgray}12.879 & 5.447 & 5.041\\\hline
							\end{tabular}\\[0.5\baselineskip]
							
							\begin{tabular}{|c|c|c||c|c||c|c||}\hline
								-E$_\mathrm{coh}^\mathrm{EXP}$=3.890 & \multicolumn{2}{c||}{DFT} & \multicolumn{2}{c||}{APE$_\mathrm{DFT}^\mathrm{PP/XC}$} & \multicolumn{2}{c||}{APE$_\mathrm{EXP}^\mathrm{PP/XC}$}\\\hline
								\diagbox{PP~~}{XC}
								& PBE & SCAN & PBE & SCAN & PBE & SCAN\\\hline
								PBE & \textbf{4.425} & 4.870 & 0 & 5.371 & 13.752 & 25.200\\
								SCAN & 4.748 & \textbf{5.147} & \cellcolor{lightgray}7.309 & 0 & 22.066 & \cellcolor{lightgray}32.307\\\hline
							\end{tabular}\\[1.5\baselineskip]	
							
							\comm{\begin{tabular}{|c|c|c|c||c|c|c||}\hline
									C$_\mathrm{11}$ & \multicolumn{3}{c||}{DFT} & \multicolumn{3}{c||}{APE$_\mathrm{DFT}^\mathrm{PP/XC}$}\\\hline
									\backslashbox{PP~~}{XC}
									& LDA & PBE & PBEsol & LDA & PBE & PBEsol\\\hline
									LDA & \textbf{116.352} & 97.294 & 106.860 & 0 & 1.808 & 2.319\\
									PBE & 119.001 & \textbf{99.086} & 109.111 & 2.277 & 0 & 0.260\\
									PBEsol & 119.212 & 99.086 & \textbf{109.396} & \cellcolor{lightgray}2.458 & 0.001 & 0\\\hline
								\end{tabular}\\[0.5\baselineskip]
								
								\begin{tabular}{|c|c|c||c|c||}\hline
									C$_\mathrm{11}$ & \multicolumn{2}{c||}{DFT} & \multicolumn{2}{c||}{APE$_\mathrm{DFT}^\mathrm{PP/XC}$}\\\hline
									\diagbox{PP~~}{XC}
									& PBE & SCAN & PBE & SCAN\\\hline
									PBE & \textbf{105.443} & 120.875 & 0 & 11.394\\
									SCAN & 125.389 & 136.419 & \cellcolor{lightgray}18.916 & 0\\\hline
								\end{tabular}\\[1.5\baselineskip]
							}	
							\begin{tabular}{|c|c|c|c||c|c|c||c|c|c||c|c||}\hline
								C$_\mathrm{11}$ & \multicolumn{3}{c||}{DFT} & \multicolumn{3}{c||}{APE$_\mathrm{DFT}^\mathrm{PP/XC}$} & C$_\mathrm{11}$ & \multicolumn{2}{c||}{DFT}& \multicolumn{2}{c||}{APE$_\mathrm{DFT}^\mathrm{PP/XC}$}\\\hline
								\backslashbox{PP}{XC} & LDA & PBE & PBEsol & LDA & PBE & PBEsol & \backslashbox{PP}{XC} & PBE & SCAN & PBE & SCAN\\\hline
								LDA & \textbf{116.352} & 97.294 & 106.860 & 0 & 1.808 & 2.319 & PBE & \textbf{105.443} & 120.875 & 0 & 11.394\\
								PBE & 119.001 & \textbf{99.086} & 109.111 & 2.277 & 0 & 0.260 & SCAN & 125.389 & 136.419 & \cellcolor{lightgray}18.916 & 0\\\cline{8-12}
								PBEsol & 119.212 & 99.086 & \textbf{109.396} & \cellcolor{lightgray}2.458 & 0.001 & 0\\\cline{1-7}
							\end{tabular}\\[1.5\baselineskip]		
							
							\comm{\begin{tabular}{|c|c|c|c||c|c|c||}\hline
									C$_\mathrm{12}$ & \multicolumn{3}{c||}{DFT} & \multicolumn{3}{c||}{APE$_\mathrm{DFT}^\mathrm{PP/XC}$}\\\hline
									\backslashbox{PP~~}{XC}
									& LDA & PBE & PBEsol & LDA & PBE & PBEsol\\\hline
									LDA & \textbf{47.764} &	36.422 & 43.758 & 0 & 0.774 & 1.076\\
									PBE & 48.310 & \textbf{36.706} & 44.125 & 1.143 & 0 & 0.247\\
									PBEsol & 48.388 & 36.715 & \textbf{44.234} & \cellcolor{lightgray}1.307 & 0.023 & 0\\\hline
								\end{tabular}\\[0.5\baselineskip]
								
								\begin{tabular}{|c|c|c||c|c||}\hline
									C$_\mathrm{12}$ & \multicolumn{2}{c||}{DFT} & \multicolumn{2}{c||}{APE$_\mathrm{DFT}^\mathrm{PP/XC}$}\\\hline
									\diagbox{PP~~}{XC}
									& PBE & SCAN & PBE & SCAN\\\hline
									PBE & \textbf{35.711} & 39.304 & 0 & 7.131\\
									SCAN & 40.887 &	\textbf{42.322}	& \cellcolor{lightgray}14.493 & 0\\\hline
								\end{tabular}\\[1.5\baselineskip]
							}
							\begin{tabular}{|c|c|c|c||c|c|c||c|c|c||c|c||}\hline
								C$_\mathrm{12}$ & \multicolumn{3}{c||}{DFT} & \multicolumn{3}{c||}{APE$_\mathrm{DFT}^\mathrm{PP/XC}$} & C$_\mathrm{12}$ & \multicolumn{2}{c||}{DFT}& \multicolumn{2}{c||}{APE$_\mathrm{DFT}^\mathrm{PP/XC}$}\\\hline
								\backslashbox{PP}{XC} & LDA & PBE & PBEsol & LDA & PBE & PBEsol & \backslashbox{PP}{XC} & PBE & SCAN & PBE & SCAN\\\hline
								LDA & \textbf{47.764} &	36.422 & 43.758 & 0 & 0.774 & 1.076 & PBE & \textbf{35.711} & 39.304 & 0 & 7.131\\
								PBE & 48.310 & \textbf{36.706} & 44.125 & 1.143 & 0 & 0.247 & SCAN & 40.887 &	\textbf{42.322}	& \cellcolor{lightgray}14.493 & 0\\\cline{8-12}
								PBEsol & 48.388 & 36.715 & \textbf{44.234} & \cellcolor{lightgray}1.307 & 0.023 & 0 \\\cline{1-7}
							\end{tabular}\\[1.5\baselineskip]		
							
							\comm{\begin{tabular}{|c|c|c|c||c|c|c||}\hline
									C$_\mathrm{44}$ & \multicolumn{3}{c||}{DFT} & \multicolumn{3}{c||}{APE$_\mathrm{DFT}^\mathrm{PP/XC}$}\\\hline
									\backslashbox{PP~~}{XC}
									& LDA & PBE & PBEsol & LDA & PBE & PBEsol\\\hline
									LDA & \textbf{56.866} & 48.723 & 51.973 & 0 & 2.445 & 2.994\\
									PBE & 58.588 & \textbf{49.944} & 53.459 & 3.028 & 0 & 0.222\\
									PBEsol & 58.701 & 50.011 & \textbf{53.578} & \cellcolor{lightgray}3.227 & 0.135 & 0\\\hline
								\end{tabular}\\[0.5\baselineskip]
								
								\begin{tabular}{|c|c|c||c|c||}\hline
									C$_\mathrm{44}$ & \multicolumn{2}{c||}{DFT} & \multicolumn{2}{c||}{APE$_\mathrm{DFT}^\mathrm{PP/XC}$}\\\hline
									\diagbox{PP~~}{XC}
									& PBE & SCAN & PBE & SCAN\\\hline
									PBE & \textbf{55.549} & 59.950 & 0 & 15.675\\
									SCAN & 65.895 & \textbf{71.095} & \cellcolor{lightgray}18.625 & 0\\\hline
								\end{tabular}\\[1.5\baselineskip]
							}
							\begin{tabular}{|c|c|c|c||c|c|c||c|c|c||c|c||}\hline
								C$_\mathrm{44}$ & \multicolumn{3}{c||}{DFT} & \multicolumn{3}{c||}{APE$_\mathrm{DFT}^\mathrm{PP/XC}$} & C$_\mathrm{44}$ & \multicolumn{2}{c||}{DFT}& \multicolumn{2}{c||}{APE$_\mathrm{DFT}^\mathrm{PP/XC}$}\\\hline
								\backslashbox{PP}{XC} & LDA & PBE & PBEsol & LDA & PBE & PBEsol & \backslashbox{PP}{XC} & PBE & SCAN & PBE & SCAN\\\hline
								LDA & \textbf{56.866} & 48.723 & 51.973 & 0 & 2.445 & 2.994 & PBE & \textbf{55.549} & 59.950 & 0 & 15.675\\
								PBE & 58.588 & \textbf{49.944} & 53.459 & 3.028 & 0 & 0.222 & SCAN & 65.895 & \textbf{71.095} & \cellcolor{lightgray}18.625 & 0\\\cline{8-12}
								PBEsol & 58.701 & 50.011 & \textbf{53.578} & \cellcolor{lightgray}3.227 & 0.135 & 0 \\\cline{1-7}
							\end{tabular}\\[1.5\baselineskip]
							{\begin{tabular}{|c|c|c|c||c|c|c||c|c|c||c|c||}\hline
									\textbf{$\gamma$} & \multicolumn{3}{c||}{DFT} & \multicolumn{3}{c||}{APE$_\mathrm{DFT}^\mathrm{PP/XC}$} & \textbf{$\gamma$} & \multicolumn{2}{c||}{DFT}& \multicolumn{2}{c||}{APE$_\mathrm{DFT}^\mathrm{PP/XC}$}\\\hline
									\backslashbox{PP}{XC} & LDA & PBE & PBEsol & LDA & PBE & PBEsol & \backslashbox{PP}{XC} & PBE & SCAN & PBE & SCAN\\\hline
									LDA & \textbf{1.742} & 1.387 & 1.600 & 0 & \cellcolor{lightgray}1.293 & 1.128 & PBE & \textbf{1.376} & 1.543 & 0 & 15.947\\
									PBE & 1.762 & \textbf{1.405} & 1.620 & 1.164 & 0 & 0.127 & SCAN & 1.641 & \textbf{1.836} & \cellcolor{lightgray}19.240 & 0\\\cline{8-12}
									PBEsol & 1.760 & 1.403 & \textbf{1.618} & 1.044 & 0.139 & 0\\\cline{1-7}
								\end{tabular}\\[1.5\baselineskip]}
							
							\begin{tabular}{|c|c|c|c||c|c|c||c|c|c||}\hline
								\textbf{B}$^\mathrm{EXP}$=79.4 & \multicolumn{3}{c||}{DFT} & \multicolumn{3}{c||}{APE$_\mathrm{DFT}^\mathrm{PP/XC}$} & \multicolumn{3}{c||}{APE$_\mathrm{EXP}^\mathrm{PP/XC}$}\\\hline
								\backslashbox{PP~~}{XC}
								& LDA & PBE & PBEsol & LDA & PBE & PBEsol & LDA & PBE & PBEsol\\\hline
								LDA & \textbf{70.627} &	56.713 & 64.792 & 0 & 1.368 & 1.763 & 11.050 & \cellcolor{lightgray}28.573 & 18.398\\
								PBE & 71.874 & \textbf{57.500} & 65.787 & 1.766 & 0 & 0.255 & 9.479 & 27.582 & 17.145\\
								PBEsol & 71.996 & 57.505 & \textbf{65.955} & \cellcolor{lightgray}1.939 & 0.010 & 0 & 9.325 & 27.575 & 16.933\\\hline
							\end{tabular}\\[0.5\baselineskip]
							
							\begin{tabular}{|c|c|c||c|c||c|c||}\hline
								\textbf{B}$^\mathrm{EXP}$=79.4 & \multicolumn{2}{c||}{DFT} & \multicolumn{2}{c||}{APE$_\mathrm{DFT}^\mathrm{PP/XC}$} & \multicolumn{2}{c||}{APE$_\mathrm{EXP}^\mathrm{PP/XC}$}\\\hline
								\diagbox{PP~~}{XC}
								& PBE & SCAN & PBE & SCAN & PBE & SCAN\\\hline
								PBE & \textbf{58.955} & 66.494 & 0 & 9.762 & \cellcolor{lightgray}25.749 & 16.254\\
								SCAN & 69.054 & \textbf{73.688} & \cellcolor{lightgray}17.130 & 0 & 13.030 & 7.195\\\hline
							\end{tabular}\\[1.5\baselineskip]
							
							\begin{tabular}{|c|c|c|c||c|c|c||}\hline
								\textbf{$\Delta_i$} & \multicolumn{3}{c||}{\textbf{$\Delta_i$}(a,b)} & \multicolumn{3}{c||}{\textbf{$\Delta_i$}(a,b)}\\\hline
								\backslashbox{PP~~}{XC} & LDA & PBE & PBEsol & \backslashbox{PP~~}{XC} & PBE & SCAN\\\hline
								LDA & 0 & 3.229 & 4.190 & PBE & 0 & \cellcolor{lightgray}28.457\\
								PBE & 4.180 & 0 & 0.301 & SCAN & 25.729 & 0 \\\cline{5-7}
								PBEsol & \cellcolor{lightgray}4.498 & 0.328 & 0 \\\cline{1-4}
							\end{tabular}
						\end{table}
						\begin{table}
							\caption{NaCl: lattice constant a$_0$\,(\AA), cohesive energy E$_\mathrm{coh}$\,(eV/atom), elastic constants C$_\mathrm{ij}$\,(GPa), surface energy \textbf{$\gamma$}\,(J/m$^2$), bulk modulus \textbf{B}\,(GPa), absolute percentage error APE$_\mathrm{DFT}^\mathrm{PP/XC}$\,(Eq.\ref{eq:mapepp}) and absolute percentage error APE$_\mathrm{EXP}^\mathrm{PP/XC}$\,(Eq.\ref{eq:mapee}) {and $\Delta_i$ gauge\,(meV/atom)\,(Eq.\ref{eq:deltapp})} (The maximum APEs {and $\Delta_i$s} are marked in gray)} \label{tab:NaCl}
							\centering
							\tiny
								\begin{tabular}{|c|c|c|c||c|c|c||c|c|c||}\hline
									a$_0^\mathrm{EXP}$=5.565 & \multicolumn{3}{c||}{DFT} & \multicolumn{3}{c||}{APE$_\mathrm{DFT}^\mathrm{PP/XC}$} & \multicolumn{3}{c||}{APE$_\mathrm{EXP}^\mathrm{PP/XC}$}\\\hline
									\backslashbox{PP~~}{XC}
									& LDA & PBE & PBEsol & LDA & PBE & PBEsol & LDA & PBE & PBEsol\\\hline
									LDA & \textbf{5.466} & 5.692 & 5.599 & 0 & 0.105 & 0.089 & 1.774 & 2.277 & 0.613\\
									PBE & 5.473 & \textbf{5.698} & 5.605 & \cellcolor{lightgray}0.115 & 0 & 0.023 & 1.662 & \cellcolor{lightgray}2.384 & 0.726\\
									PBEsol & 5.471 & 5.696 & \textbf{5.604} & 0.089 & 0.026 & 0 & 1.687 & 2.357 & 0.703\\\hline
								\end{tabular}\\[0.5\baselineskip]
								
								\begin{tabular}{|c|c|c||c|c||c|c||}\hline
									a$_0^\mathrm{EXP}$=5.565 & \multicolumn{2}{c||}{DFT} & \multicolumn{2}{c||}{APE$_\mathrm{DFT}^\mathrm{PP/XC}$} & \multicolumn{2}{c||}{APE$_\mathrm{EXP}^\mathrm{PP/XC}$}\\\hline
									\diagbox{PP~~}{XC}
									& PBE & SCAN & PBE & SCAN & PBE & SCAN\\\hline
									PBE & \textbf{5.664} & 5.585 & 0 & 1.689 & \cellcolor{lightgray}1.772 & 0.352\\
									SCAN & 5.565 & \textbf{5.492} &	\cellcolor{lightgray}1.738 & 0 & 0.003 & 1.316\\\hline
								\end{tabular}\\[1.5\baselineskip]
								
								\begin{tabular}{|c|c|c|c||c|c|c||c|c|c||}\hline
									-E$_\mathrm{coh}^\mathrm{EXP}$=3.340 & \multicolumn{3}{c||}{DFT} & \multicolumn{3}{c||}{APE$_\mathrm{DFT}^\mathrm{PP/XC}$} & \multicolumn{3}{c||}{APE$_\mathrm{EXP}^\mathrm{PP/XC}$}\\\hline
									\backslashbox{PP~~}{XC}
									& LDA & PBE & PBEsol & LDA & PBE & PBEsol & LDA & PBE & PBEsol\\\hline
									LDA & \textbf{3.249} & 2.931 & 3.036 & 0 & 0.110 & 0.364 & 2.722 & 12.255 & 9.089\\
									PBE & 3.216 & \textbf{2.934} & 3.022 & \cellcolor{lightgray}1.026 &	0 & 0.123 & 3.720 & 12.159 & 9.530\\
									PBEsol & 3.220 & 2.919 & \textbf{3.025} & 0.904 & 0.507 & 0 & 3.602 & \cellcolor{lightgray}12.604 & 9.419\\\hline
								\end{tabular}\\[0.5\baselineskip]
								
								\begin{tabular}{|c|c|c||c|c||c|c||}\hline
									-E$_\mathrm{coh}^\mathrm{EXP}$=3.340 & \multicolumn{2}{c||}{DFT} & \multicolumn{2}{c||}{APE$_\mathrm{DFT}^\mathrm{PP/XC}$} & \multicolumn{2}{c||}{APE$_\mathrm{EXP}^\mathrm{PP/XC}$}\\\hline
									\diagbox{PP~~}{XC}
									& PBE & SCAN & PBE & SCAN & PBE & SCAN\\\hline
									PBE & \textbf{3.382} & 3.668 & 0 & \cellcolor{lightgray}2.316 & 1.254 & 9.831\\
									SCAN & 3.451 & \textbf{3.755} & 2.050 &	0 & 3.329 & \cellcolor{lightgray}12.435\\\hline
								\end{tabular}\\[1.5\baselineskip]	
								
								\comm{\begin{tabular}{|c|c|c|c||c|c|c||}\hline
										C$_\mathrm{11}$ & \multicolumn{3}{c||}{DFT} & \multicolumn{3}{c||}{APE$_\mathrm{DFT}^\mathrm{PP/XC}$}\\\hline
										\backslashbox{PP~~}{XC}
										& LDA & PBE & PBEsol & LDA & PBE & PBEsol\\\hline
										LDA & 70.206 & 47.260 & 54.876 & 0 & 0.025 & 0.192\\
										PBE & 69.925 & \textbf{47.249} & 54.736 & 0.400 & 0 & 0.064\\
										PBEsol & 69.996 & 47.582 & 54.771 & 0.299 & \cellcolor{lightgray}0.706 & 0\\\hline
									\end{tabular}\\[0.5\baselineskip]
									
									\begin{tabular}{|c|c|c||c|c||}\hline
										C$_\mathrm{11}$ & \multicolumn{2}{c||}{DFT} & \multicolumn{2}{c||}{APE$_\mathrm{DFT}^\mathrm{PP/XC}$}\\\hline
										\diagbox{PP~~}{XC}
										& PBE & SCAN & PBE & SCAN\\\hline
										PBE	& \textbf{43.026} &	54.160 & 0 & \cellcolor{lightgray}5.895\\
										SCAN & 44.921 & \textbf{57.552} & 4.403 & 0\\\hline
									\end{tabular}\\[1.5\baselineskip]
								}
								\begin{tabular}{|c|c|c|c||c|c|c||c|c|c||c|c||}\hline
									C$_\mathrm{11}$ & \multicolumn{3}{c||}{DFT} & \multicolumn{3}{c||}{APE$_\mathrm{DFT}^\mathrm{PP/XC}$} & C$_\mathrm{11}$ & \multicolumn{2}{c||}{DFT}& \multicolumn{2}{c||}{APE$_\mathrm{DFT}^\mathrm{PP/XC}$}\\\hline
									\backslashbox{PP}{XC} & LDA & PBE & PBEsol & LDA & PBE & PBEsol & \backslashbox{PP}{XC} & PBE & SCAN & PBE & SCAN\\\hline
									LDA & 70.206 & 47.260 & 54.876 & 0 & 0.025 & 0.192 & PBE & \textbf{43.026} &	54.160 & 0 & \cellcolor{lightgray}5.895\\
									PBE & 69.925 & \textbf{47.249} & 54.736 & 0.400 & 0 & 0.064 & SCAN & 44.921 & \textbf{57.552} & 4.403 & 0\\\cline{8-12}
									PBEsol & 69.996 & 47.582 & 54.771 & 0.299 & \cellcolor{lightgray}0.706 & 0\\\cline{1-7}
								\end{tabular}\\[1.5\baselineskip]
								
								\comm{m\begin{tabular}{|c|c|c|c||c|c|c||}\hline
										C$_\mathrm{12}$ & \multicolumn{3}{c||}{DFT} & \multicolumn{3}{c||}{APE$_\mathrm{DFT}^\mathrm{PP/XC}$}\\\hline
										\backslashbox{PP~~}{XC}
										& LDA & PBE & PBEsol & LDA & PBE & PBEsol\\\hline
										LDA & \textbf{13.055} & 11.839 & 11.339 & 0 & 0.381 & 0.361\\
										PBE & 13.063 & \textbf{11.884} & 11.368 & 0.060 & 0 & 0.109\\
										PBEsol & 13.060 & 12.021 & \textbf{11.380} & 0.035 & \cellcolor{lightgray}1.147 & 0\\\hline
									\end{tabular}\\[0.5\baselineskip]
									
									\begin{tabular}{|c|c|c||c|c||}\hline
										C$_\mathrm{12}$ & \multicolumn{2}{c||}{DFT} & \multicolumn{2}{c||}{APE$_\mathrm{DFT}^\mathrm{PP/XC}$}\\\hline
										\diagbox{PP~~}{XC}
										& PBE & SCAN & PBE & SCAN\\\hline
										PBE & \textbf{9.763} & 10.608 & 0 & \cellcolor{lightgray}10.614\\
										SCAN & 10.465 & \textbf{11.868} & 7.184 & 0\\\hline
									\end{tabular}\\[1.5\baselineskip]
								}
								\begin{tabular}{|c|c|c|c||c|c|c||c|c|c||c|c||}\hline
									C$_\mathrm{12}$ & \multicolumn{3}{c||}{DFT} & \multicolumn{3}{c||}{APE$_\mathrm{DFT}^\mathrm{PP/XC}$} & C$_\mathrm{12}$ & \multicolumn{2}{c||}{DFT}& \multicolumn{2}{c||}{APE$_\mathrm{DFT}^\mathrm{PP/XC}$}\\\hline
									\backslashbox{PP}{XC} & LDA & PBE & PBEsol & LDA & PBE & PBEsol & \backslashbox{PP}{XC} & PBE & SCAN & PBE & SCAN\\\hline
									LDA & \textbf{13.055} & 11.839 & 11.339 & 0 & 0.381 & 0.361 & PBE & \textbf{9.763} & 10.608 & 0 & \cellcolor{lightgray}10.614\\
									PBE & 13.063 & \textbf{11.884} & 11.368 & 0.060 & 0 & 0.109 & SCAN & 10.465 & \textbf{11.868} & 7.184 & 0\\\cline{8-12}
									PBEsol & 13.060 & 12.021 & \textbf{11.380} & 0.035 & \cellcolor{lightgray}1.147 & 0\\\cline{1-7}
								\end{tabular}\\[1.5\baselineskip]
								
								\comm{\begin{tabular}{|c|c|c|c||c|c|c||}\hline
										C$_\mathrm{44}$ & \multicolumn{3}{c||}{DFT} & \multicolumn{3}{c||}{APE$_\mathrm{DFT}^\mathrm{PP/XC}$}\\\hline
										\backslashbox{PP~~}{XC}
										& LDA & PBE & PBEsol & LDA & PBE & PBEsol\\\hline
										LDA	& \textbf{12.896} & 12.402 & 12.071 & 0 & 0.369 & 0.251\\
										PBE & 12.774 & \textbf{12.356} & 11.998 & 0.950 & 0 & 0.355\\
										PBEsol & 12.813 & 12.517 & \textbf{12.041} & 0.650 & \cellcolor{lightgray}1.301 & 0\\\hline
									\end{tabular}\\[0.5\baselineskip]
									
									\begin{tabular}{|c|c|c||c|c||}\hline
										C$_\mathrm{44}$ & \multicolumn{2}{c||}{DFT} & \multicolumn{2}{c||}{APE$_\mathrm{DFT}^\mathrm{PP/XC}$}\\\hline
										\diagbox{PP~~}{XC}
										& PBE & SCAN & PBE & SCAN\\\hline
										PBE	& \textbf{13.460} & 14.204 & 0 & \cellcolor{lightgray}14.136\\
										SCAN & 15.125 & \textbf{16.542} & 12.366 & 0\\\hline
									\end{tabular}\\[1.5\baselineskip]
								}
								\begin{tabular}{|c|c|c|c||c|c|c||c|c|c||c|c||}\hline
									C$_\mathrm{44}$ & \multicolumn{3}{c||}{DFT} & \multicolumn{3}{c||}{APE$_\mathrm{DFT}^\mathrm{PP/XC}$} & C$_\mathrm{44}$ & \multicolumn{2}{c||}{DFT}& \multicolumn{2}{c||}{APE$_\mathrm{DFT}^\mathrm{PP/XC}$}\\\hline
									\backslashbox{PP}{XC} & LDA & PBE & PBEsol & LDA & PBE & PBEsol & \backslashbox{PP}{XC} & PBE & SCAN & PBE & SCAN\\\hline
									LDA & \textbf{12.896} & 12.402 & 12.071 & 0 & 0.369 & 0.251 & PBE & \textbf{13.460} & 14.204 & 0 & \cellcolor{lightgray}14.136\\
									PBE & 12.774 & \textbf{12.356} & 11.998 & 0.950 & 0 & 0.355 & SCAN & 15.125 & \textbf{16.542} & 12.366 & 0\\\cline{8-12}
									PBEsol & 12.813 & 12.517 & \textbf{12.041} & 0.650 & 1.301 & 0\\\cline{1-7}
								\end{tabular}\\[1.5\baselineskip]
								{\begin{tabular}{|c|c|c|c||c|c|c||c|c|c||c|c||}\hline
										\textbf{$\gamma$} & \multicolumn{3}{c||}{DFT} & \multicolumn{3}{c||}{APE$_\mathrm{DFT}^\mathrm{PP/XC}$} & \textbf{$\gamma$} & \multicolumn{2}{c||}{DFT}& \multicolumn{2}{c||}{APE$_\mathrm{DFT}^\mathrm{PP/XC}$}\\\hline
										\backslashbox{PP}{XC} & LDA & PBE & PBEsol & LDA & PBE & PBEsol & \backslashbox{PP}{XC} & PBE & SCAN & PBE & SCAN\\\hline
										LDA & \textbf{0.244} & 0.149 & 0.173 & 0 & 0.126 & \cellcolor{lightgray}0.145 & PBE & \textbf{0.166} & 0.211 & 0 & \cellcolor{lightgray}4.964\\
										PBE & 0.244 & \textbf{0.149} & 0.173 & 0.047 & 0 & 0.105 & SCAN & 0.173 & \textbf{0.222} & 3.973 & 0\\\cline{8-12}
										PBEsol & 0.244 & 0.149 & \textbf{0.173} & 0.107 & 0.046 & 0\\\cline{1-7}
									\end{tabular}\\[1.5\baselineskip]}
								
								\begin{tabular}{|c|c|c|c||c|c|c||c|c|c||}\hline
									\textbf{B}$^\mathrm{EXP}$=27.6 & \multicolumn{3}{c||}{DFT} & \multicolumn{3}{c||}{APE$_\mathrm{DFT}^\mathrm{PP/XC}$} & \multicolumn{3}{c||}{APE$_\mathrm{EXP}^\mathrm{PP/XC}$}\\\hline
									\backslashbox{PP~~}{XC}
									& LDA & PBE & PBEsol & LDA & PBE & PBEsol & LDA & PBE & PBEsol\\\hline
									LDA & \textbf{32.105} & 23.646 & 25.851 & 0 & 0.111 & 0.023 & \cellcolor{lightgray}16.324 & 14.325 & 6.336\\
									PBE & 32.017 & \textbf{23.673} & 25.824 & 0.275 & 0 & 0.077 & 16.004 & 14.230 & 6.436\\
									PBEsol & 32.038 & 23.875 & \textbf{25.844} & 0.209 & \cellcolor{lightgray}0.854 & 0 & 16.081 & 13.498 & 6.364\\\hline
								\end{tabular}\\[0.5\baselineskip]
								
								\begin{tabular}{|c|c|c||c|c||c|c||}\hline
									\textbf{B}$^\mathrm{EXP}$=27.6 & \multicolumn{2}{c||}{DFT} & \multicolumn{2}{c||}{APE$_\mathrm{DFT}^\mathrm{PP/XC}$} & \multicolumn{2}{c||}{APE$_\mathrm{EXP}^\mathrm{PP/XC}$}\\\hline
									\diagbox{PP~~}{XC}
									& PBE & SCAN & PBE & SCAN & PBE & SCAN\\\hline
									PBE & \textbf{20.851} & 25.126 & 0 & \cellcolor{lightgray}7.273 & \cellcolor{lightgray}24.453 & 8.965\\
									SCAN & 21.950 & \textbf{27.096} & 5.271 & 0 & 20.471 & 1.825\\\hline
								\end{tabular}\\[1.5\baselineskip]
								
								\begin{tabular}{|c|c|c|c||c|c|c||}\hline
									\textbf{$\Delta_i$} & \multicolumn{3}{c||}{\textbf{$\Delta_i$}(a,b)} & \multicolumn{3}{c||}{\textbf{$\Delta_i$}(a,b)}\\\hline
									\backslashbox{PP~~}{XC} & LDA & PBE & PBEsol & \backslashbox{PP~~}{XC} & PBE & SCAN\\\hline
									LDA & 0 & 0.376 & 0.330  & PBE & 0 & \cellcolor{lightgray}6.044\\
									PBE & \cellcolor{lightgray}0.491 & 0 & 0.086 & SCAN & 5.414 & 0 \\\cline{5-7}
									PBEsol & 0.381 & 0.093 & 0 \\\cline{1-4}
								\end{tabular}
							\end{table}
							\begin{table}
								\caption{LiCl: lattice constant a$_0$\,(\AA), cohesive energy E$_\mathrm{coh}$\,(eV/atom), elastic constants C$_\mathrm{ij}$\,(GPa), surface energy \textbf{$\gamma$}\,(J/m$^2$), bulk modulus \textbf{B}\,(GPa), absolute percentage error APE$_\mathrm{DFT}^\mathrm{PP/XC}$\,(Eq.\ref{eq:mapepp}) and absolute percentage error APE$_\mathrm{EXP}^\mathrm{PP/XC}$\,(Eq.\ref{eq:mapee}) {and $\Delta_i$ gauge\,(meV/atom)\,(Eq.\ref{eq:deltapp})} (The maximum APEs {and $\Delta_i$s} are marked in gray)} \label{tab:LiCl}
								\centering
								\tiny
									\begin{tabular}{|c|c|c|c||c|c|c||c|c|c||}\hline
										a$_0^\mathrm{EXP}$=5.07 & \multicolumn{3}{c||}{DFT} & \multicolumn{3}{c||}{APE$_\mathrm{DFT}^\mathrm{PP/XC}$} & \multicolumn{3}{c||}{APE$_\mathrm{EXP}^\mathrm{PP/XC}$}\\\hline
										\backslashbox{PP~~}{XC}
										& LDA & PBE & PBEsol & LDA & PBE & PBEsol & LDA & PBE & PBEsol\\\hline
										LDA & \textbf{4.965} & 5.141 & 5.057 & 0 & 0.174 & 0.124 & \cellcolor{lightgray}2.081 & 1.401 & 0.251\\
										PBE & 4.974 & \textbf{5.150} & 5.066 & \cellcolor{lightgray}0.199 & 0 & 0.058 & 1.886 & 1.578 & 0.070\\
										PBEsol & 4.971 & 5.147 & \textbf{5.064} & 0.132 & 0.049 & 0 & 1.952 & 1.528 & 0.128\\\hline
									\end{tabular}\\[0.5\baselineskip]
									
									\begin{tabular}{|c|c|c||c|c||c|c||}\hline
										a$_0^\mathrm{EXP}$=5.07 & \multicolumn{2}{c||}{DFT} & \multicolumn{2}{c||}{APE$_\mathrm{DFT}^\mathrm{PP/XC}$} & \multicolumn{2}{c||}{APE$_\mathrm{EXP}^\mathrm{PP/XC}$}\\\hline
										\diagbox{PP~~}{XC}
										& PBE & SCAN & PBE & SCAN & PBE & SCAN\\\hline
										PBE & \textbf{5.170} & 5.105 & 0 & \cellcolor{lightgray}0.100 & \cellcolor{lightgray}1.968 & 0.689\\
										SCAN & 5.165 & \textbf{5.100} & 0.093 & 0 & 1.874 & 0.589\\\hline
									\end{tabular}\\[1.5\baselineskip]
									
									\begin{tabular}{|c|c|c|c||c|c|c||c|c|c||}\hline
										-E$_\mathrm{coh}^\mathrm{EXP}$=3.586 & \multicolumn{3}{c||}{DFT} & \multicolumn{3}{c||}{APE$_\mathrm{DFT}^\mathrm{PP/XC}$} & \multicolumn{3}{c||}{APE$_\mathrm{EXP}^\mathrm{PP/XC}$}\\\hline
										\backslashbox{PP~~}{XC}
										& LDA & PBE & PBEsol & LDA & PBE & PBEsol & LDA & PBE & PBEsol\\\hline
										LDA & \textbf{3.601} & 3.248 & 3.366 & 0 & 0.567 & 0.226 & 0.420 & 9.439 & 6.145\\
										PBE & 3.580 & \textbf{3.229} & 3.355 & \cellcolor{lightgray}0.578 & 0 & 0.093 & 0.161 & \cellcolor{lightgray}9.950 & 6.444\\
										PBEsol & 3.584 & 3.232 & \textbf{3.358} & 0.474 & 0.088 & 0 & 0.056 & 9.871 & 6.357\\\hline
									\end{tabular}\\[0.5\baselineskip]
									
									\begin{tabular}{|c|c|c||c|c||c|c||}\hline
										-E$_\mathrm{coh}^\mathrm{EXP}$=3.586 & \multicolumn{2}{c||}{DFT} & \multicolumn{2}{c||}{APE$_\mathrm{DFT}^\mathrm{PP/XC}$} & \multicolumn{2}{c||}{APE$_\mathrm{EXP}^\mathrm{PP/XC}$}\\\hline
										\diagbox{PP~~}{XC}
										& PBE & SCAN & PBE & SCAN & PBE & SCAN\\\hline
										PBE & \textbf{3.640} & 3.987 & 0 & \cellcolor{lightgray}0.979 & 1.496 & 11.192\\
										SCAN & 3.650 & \textbf{4.027} & 0.294 & 0 & 1.794 & \cellcolor{lightgray}12.291\\\hline
									\end{tabular}\\[1.5\baselineskip]	
									
									\comm{\begin{tabular}{|c|c|c|c||c|c|c||}\hline
											C$_\mathrm{11}$ & \multicolumn{3}{c||}{DFT} & \multicolumn{3}{c||}{APE$_\mathrm{DFT}^\mathrm{PP/XC}$}\\\hline
											\backslashbox{PP~~}{XC}
											& LDA & PBE & PBEsol & LDA & PBE & PBEsol\\\hline
											LDA & \textbf{76.310} & 52.098 & 63.533 & 0 & 0.567 & 0.203\\
											PBE & 75.863 & \textbf{51.804} & 63.335 & \cellcolor{lightgray}0.586 & 0 & 0.109\\
											PBEsol & 76.011 & 51.799 & \textbf{63.404} & 0.391 & 0.010 & 0\\\hline
										\end{tabular}\\[0.5\baselineskip]
										
										\begin{tabular}{|c|c|c||c|c||}\hline
											C$_\mathrm{11}$ & \multicolumn{2}{c||}{DFT} & \multicolumn{2}{c||}{APE$_\mathrm{DFT}^\mathrm{PP/XC}$}\\\hline
											\diagbox{PP~~}{XC}
											& PBE & SCAN & PBE & SCAN\\\hline
											PBE & \textbf{75.420} & 77.310 & 0 & \cellcolor{lightgray}0.300\\
											SCAN & 75.201 & \textbf{77.540} & 0.290 & 0\\\hline
										\end{tabular}\\[1.5\baselineskip]
									}	
									\begin{tabular}{|c|c|c|c||c|c|c||c|c|c||c|c||}\hline
										C$_\mathrm{11}$ & \multicolumn{3}{c||}{DFT} & \multicolumn{3}{c||}{APE$_\mathrm{DFT}^\mathrm{PP/XC}$} & C$_\mathrm{11}$ & \multicolumn{2}{c||}{DFT}& \multicolumn{2}{c||}{APE$_\mathrm{DFT}^\mathrm{PP/XC}$}\\\hline
										\backslashbox{PP}{XC} & LDA & PBE & PBEsol & LDA & PBE & PBEsol & \backslashbox{PP}{XC} & PBE & SCAN & PBE & SCAN\\\hline
										LDA & \textbf{76.310} & 52.098 & 63.533 & 0 & 0.567 & 0.203 & PBE & \textbf{75.420} & 77.310 & 0 & \cellcolor{lightgray}0.300\\
										PBE & 75.863 & \textbf{51.804} & 63.335 & \cellcolor{lightgray}0.586 & 0 & 0.109 & SCAN & 75.201 & \textbf{77.540} & 0.290 & 0\\\cline{8-12}
										PBEsol & 76.011 & 51.799 & \textbf{63.404} & 0.391 & 0.010 & 0\\\cline{1-7}
									\end{tabular}\\[1.5\baselineskip]
									
									\comm{\begin{tabular}{|c|c|c|c||c|c|c||}\hline
											C$_\mathrm{12}$ & \multicolumn{3}{c||}{DFT} & \multicolumn{3}{c||}{APE$_\mathrm{DFT}^\mathrm{PP/XC}$}\\\hline
											\backslashbox{PP~~}{XC}
											& LDA & PBE & PBEsol & LDA & PBE & PBEsol\\\hline
											LDA & \textbf{23.150} & 21.958 & 21.010 & 0 & \cellcolor{lightgray}0.854 & 0.263\\
											PBE & 22.969 & \textbf{21.772} & 20.896 & 0.782 & 0 & 0.284\\
											PBEsol & 23.050 & 21.818 & \textbf{20.955} & 0.432 & 0.211 & 0\\\hline
										\end{tabular}\\[0.5\baselineskip]
										
										\begin{tabular}{|c|c|c||c|c||}\hline
											C$_\mathrm{12}$ & \multicolumn{2}{c||}{DFT} & \multicolumn{2}{c||}{APE$_\mathrm{DFT}^\mathrm{PP/XC}$}\\\hline
											\diagbox{PP~~}{XC}
											& PBE & SCAN & PBE & SCAN\\\hline
											PBE & \textbf{19.585} & 20.223 & 0 & 0.435\\
											SCAN & 19.720 & \textbf{20.135} & \cellcolor{lightgray}0.690 & 0\\\hline
										\end{tabular}\\[1.5\baselineskip]
									}
									\begin{tabular}{|c|c|c|c||c|c|c||c|c|c||c|c||}\hline
										C$_\mathrm{12}$ & \multicolumn{3}{c||}{DFT} & \multicolumn{3}{c||}{APE$_\mathrm{DFT}^\mathrm{PP/XC}$} & C$_\mathrm{12}$ & \multicolumn{2}{c||}{DFT}& \multicolumn{2}{c||}{APE$_\mathrm{DFT}^\mathrm{PP/XC}$}\\\hline
										\backslashbox{PP}{XC} & LDA & PBE & PBEsol & LDA & PBE & PBEsol & \backslashbox{PP}{XC} & PBE & SCAN & PBE & SCAN\\\hline
										LDA & \textbf{23.150} & 21.958 & 21.010 & 0 & \cellcolor{lightgray}0.854 & 0.263 & PBE & \textbf{19.585} & 20.223 & 0 & 0.435\\
										PBE & 22.969 & \textbf{21.772} & 20.896 & 0.782 & 0 & 0.284 & SCAN & 19.720 & \textbf{20.135} & \cellcolor{lightgray}0.690 & 0\\\cline{8-12}
										PBEsol & 23.050 & 21.818 & \textbf{20.955} & 0.432 & 0.211 & 0\\\cline{1-7}
									\end{tabular}\\[1.5\baselineskip]
									
									\comm{\begin{tabular}{|c|c|c|c||c|c|c||}\hline
											C$_\mathrm{44}$ & \multicolumn{3}{c||}{DFT} & \multicolumn{3}{c||}{APE$_\mathrm{DFT}^\mathrm{PP/XC}$}\\\hline
											\backslashbox{PP~~}{XC}
											& LDA & PBE & PBEsol & LDA & PBE & PBEsol\\\hline
											LDA & \textbf{28.170} & 25.187 & 25.618 & 0 & 1.048 & 0.445\\
											PBE & 27.839 & \textbf{24.926} & 25.397 & \cellcolor{lightgray}1.174 & 0 & 0.423\\
											PBEsol & 27.992 & 25.002 & \textbf{25.505} & 0.632 & 0.306 & 0\\\hline
										\end{tabular}\\[0.5\baselineskip]
										
										\begin{tabular}{|c|c|c||c|c||}\hline
											C$_\mathrm{44}$ & \multicolumn{2}{c||}{DFT} & \multicolumn{2}{c||}{APE$_\mathrm{DFT}^\mathrm{PP/XC}$}\\\hline
											\diagbox{PP~~}{XC}
											& PBE & SCAN & PBE & SCAN\\\hline
											PBE & \textbf{25.338} & 25.703 & 0 & \cellcolor{lightgray}1.201\\
											SCAN & 25.492 & \textbf{26.016} & 0.608 & 0\\\hline
										\end{tabular}\\[1.5\baselineskip]
									}
									\begin{tabular}{|c|c|c|c||c|c|c||c|c|c||c|c||}\hline
										C$_\mathrm{44}$ & \multicolumn{3}{c||}{DFT} & \multicolumn{3}{c||}{APE$_\mathrm{DFT}^\mathrm{PP/XC}$} & C$_\mathrm{44}$ & \multicolumn{2}{c||}{DFT}& \multicolumn{2}{c||}{APE$_\mathrm{DFT}^\mathrm{PP/XC}$}\\\hline
										\backslashbox{PP}{XC} & LDA & PBE & PBEsol & LDA & PBE & PBEsol & \backslashbox{PP}{XC} & PBE & SCAN & PBE & SCAN\\\hline
										LDA & \textbf{28.170} & 25.187 & 25.618 & 0 & 1.048 & 0.445 & PBE & \textbf{25.338} & 25.703 & 0 & \cellcolor{lightgray}1.201\\
										PBE & 27.839 & \textbf{24.926} & 25.397 & \cellcolor{lightgray}1.174 & 0 & 0.423 & SCAN & 25.492 & \textbf{26.016} & 0.608 & 0\\\cline{8-12}
										PBEsol & 27.992 & 25.002 & \textbf{25.505} & 0.632 & 0.306 & 0\\\cline{1-7}
									\end{tabular}\\[1.5\baselineskip]
									{\begin{tabular}{|c|c|c|c||c|c|c||c|c|c||c|c||}\hline
											\textbf{$\gamma$} & \multicolumn{3}{c||}{DFT} & \multicolumn{3}{c||}{APE$_\mathrm{DFT}^\mathrm{PP/XC}$} & \textbf{$\gamma$} & \multicolumn{2}{c||}{DFT}& \multicolumn{2}{c||}{APE$_\mathrm{DFT}^\mathrm{PP/XC}$}\\\hline
											\backslashbox{PP}{XC} & LDA & PBE & PBEsol & LDA & PBE & PBEsol & \backslashbox{PP}{XC} & PBE & SCAN & PBE & SCAN\\\hline
											LDA & \textbf{0.283} & 0.158 & 0.206 & 0 & \cellcolor{lightgray}1.313 & 0.590 & PBE & \textbf{0.189} & 0.200 & 0 & \cellcolor{lightgray}0.199\\
											PBE & 0.285 & \textbf{0.160} & 0.207 & 0.805 & 0 & 0.290 & SCAN & 0.189 & \textbf{0.200} & 0.052 & 0\\\cline{8-12}
											PBEsol & 0.285 & 0.160 & \textbf{0.207} & 0.719 & 0.277 & 0\\\cline{1-7}
										\end{tabular}\\[1.5\baselineskip]}
									
									\begin{tabular}{|c|c|c|c||c|c|c||c|c|c||}\hline
										\textbf{B}$^\mathrm{EXP}$=38.7 & \multicolumn{3}{c||}{DFT} & \multicolumn{3}{c||}{APE$_\mathrm{DFT}^\mathrm{PP/XC}$} & \multicolumn{3}{c||}{APE$_\mathrm{EXP}^\mathrm{PP/XC}$}\\\hline
										\backslashbox{PP~~}{XC}
										& LDA & PBE & PBEsol & LDA & PBE & PBEsol & LDA & PBE & PBEsol\\\hline
										LDA & \textbf{40.870} & 32.004 & 35.184 & 0 & \cellcolor{lightgray}0.698 & 0.227 & 5.607 & 17.302 & 9.084\\
										PBE & 40.600 & \textbf{31.782} & 35.042 & 0.660 & 0 & 0.179 & 4.910 & \cellcolor{lightgray}17.875 & 9.452\\
										PBEsol & 40.704 & 31.811 & \textbf{35.105} & 0.407 & 0.091 & 0 & 5.178 & 17.800 & 9.290\\\hline
									\end{tabular}\\[0.5\baselineskip]
									
									\begin{tabular}{|c|c|c||c|c||c|c||}\hline
										\textbf{B}$^\mathrm{EXP}$=38.7 & \multicolumn{2}{c||}{DFT} & \multicolumn{2}{c||}{APE$_\mathrm{DFT}^\mathrm{PP/XC}$} & \multicolumn{2}{c||}{APE$_\mathrm{EXP}^\mathrm{PP/XC}$}\\\hline
										\diagbox{PP~~}{XC}
										& PBE & SCAN & PBE & SCAN & PBE & SCAN\\\hline
										PBE & \textbf{38.197} & 39.252 & 0 & \cellcolor{lightgray}0.046 & 1.300 & 1.426\\
										SCAN & 38.214 & \textbf{39.270} & 0.045 & 0 & 1.256 & \cellcolor{lightgray}1.473\\\hline
									\end{tabular}\\[1.5\baselineskip]
									
									\begin{tabular}{|c|c|c|c||c|c|c||}\hline
										\textbf{$\Delta_i$} & \multicolumn{3}{c||}{\textbf{$\Delta_i$}(a,b)} & \multicolumn{3}{c||}{\textbf{$\Delta_i$}(a,b)}\\\hline
										\backslashbox{PP~~}{XC} & LDA & PBE & PBEsol & \backslashbox{PP~~}{XC} & PBE & SCAN\\\hline
										LDA & 0 & 0.618 & 0.462 & PBE & 0 & \cellcolor{lightgray}0.425\\
										PBE & \cellcolor{lightgray}0.809 & 0 & 0.214 & SCAN & 0.400 & 0 \\\cline{5-7}
										PBEsol & 0.537 & 0.173 & 0 \\\cline{1-4}
									\end{tabular}
								\end{table}
								\begin{table}  
									\caption{LiF: lattice constant a$_0$\,(\AA), cohesive energy E$_\mathrm{coh}$\,(eV/atom), elastic constants C$_\mathrm{ij}$\,(GPa), surface energy \textbf{$\gamma$}\,(J/m$^2$), bulk modulus \textbf{B}\,(GPa), absolute percentage error APE$_\mathrm{DFT}^\mathrm{PP/XC}$\,(Eq.\ref{eq:mapepp}) and absolute percentage error APE$_\mathrm{EXP}^\mathrm{PP/XC}$\,(Eq.\ref{eq:mapee}) {and $\Delta_i$ gauge\,(meV/atom)\,(Eq.\ref{eq:deltapp})} (The maximum APEs {and $\Delta_i$s} are marked in gray)} \label{tab:LiF}
									\centering
									\tiny
										\begin{tabular}{|c|c|c|c||c|c|c||c|c|c||}\hline
											a$_0^\mathrm{EXP}$=3.974 & \multicolumn{3}{c||}{DFT} & \multicolumn{3}{c||}{APE$_\mathrm{DFT}^\mathrm{PP/XC}$} & \multicolumn{3}{c||}{APE$_\mathrm{EXP}^\mathrm{PP/XC}$}\\\hline
											\backslashbox{PP~~}{XC}
											& LDA & PBE & PBEsol & LDA & PBE & PBEsol & LDA & PBE & PBEsol\\\hline
											LDA & \textbf{3.905} & 4.0487 & 3.994 & 0 & 0.316 & 0.198 & 1.727 & 1.879 & 0.491\\
											PBE	& 3.918 & \textbf{4.061} & 4.006 & \cellcolor{lightgray}0.330 & 0 & 0.119 & 1.403 & \cellcolor{lightgray}2.202 & 0.810\\
											PBEsol & 3.913 & 4.057 & \textbf{4.001} & 0.203 & 0.119 & 0 & 1.528 & 2.080 & 0.690\\\hline
										\end{tabular}\\[0.5\baselineskip]
										
										\begin{tabular}{|c|c|c||c|c||c|c||}\hline
											a$_0^\mathrm{EXP}$=3.974 & \multicolumn{2}{c||}{DFT} & \multicolumn{2}{c||}{APE$_\mathrm{DFT}^\mathrm{PP/XC}$} & \multicolumn{2}{c||}{APE$_\mathrm{EXP}^\mathrm{PP/XC}$}\\\hline
											\diagbox{PP~~}{XC}
											& PBE & SCAN & PBE & SCAN & PBE & SCAN\\\hline
											PBE & \textbf{4.087} & 4.008 & 0 & 0.089 & \cellcolor{lightgray}2.836 & 0.863\\
											SCAN & 4.083 & \textbf{4.005} & \cellcolor{lightgray}0.096 & 0 & 2.737 & 0.774\\\hline
										\end{tabular}\\[1.5\baselineskip]
										
										\begin{tabular}{|c|c|c|c||c|c|c||c|c|c||}\hline
											-E$_\mathrm{coh}^\mathrm{EXP}$=4.46 & \multicolumn{3}{c||}{DFT} & \multicolumn{3}{c||}{APE$_\mathrm{DFT}^\mathrm{PP/XC}$} & \multicolumn{3}{c||}{APE$_\mathrm{EXP}^\mathrm{PP/XC}$}\\\hline
											\backslashbox{PP~~}{XC}
											& LDA & PBE & PBEsol & LDA & PBE & PBEsol & LDA & PBE & PBEsol\\\hline
											LDA & \textbf{4.819} & 4.412 & 4.501 & 0 & \cellcolor{lightgray}0.921 & 0.475 & \cellcolor{lightgray}8.057 & 1.083 & 0.912\\
											PBE & 4.777 & \textbf{4.371} & 4.469 & 0.884 & 0 & 0.236 & 7.102 & 1.985 & 0.198\\
											PBEsol & 4.788 & 4.382 & \textbf{4.479} & 0.645 & 0.243 & 0 & 7.360 & 1.747 & 0.435\\\hline
										\end{tabular}\\[0.5\baselineskip]
										
										\begin{tabular}{|c|c|c||c|c||c|c||}\hline
											-E$_\mathrm{coh}^\mathrm{EXP}$=4.46 & \multicolumn{2}{c||}{DFT} & \multicolumn{2}{c||}{APE$_\mathrm{DFT}^\mathrm{PP/XC}$} & \multicolumn{2}{c||}{APE$_\mathrm{EXP}^\mathrm{PP/XC}$}\\\hline
											\diagbox{PP~~}{XC}
											& PBE & SCAN & PBE & SCAN & PBE & SCAN\\\hline
											PBE & \textbf{4.789} & 4.592 & 0 & \cellcolor{lightgray}0.510 & 7.368 & 2.955\\
											SCAN & 4.805 & \textbf{4.569} &	0.335 & 0 & \cellcolor{lightgray}7.728 & 2.433\\\hline
										\end{tabular}\\[1.5\baselineskip]	
										
										\comm{\begin{tabular}{|c|c|c|c||c|c|c||}\hline
												C$_\mathrm{11}$ & \multicolumn{3}{c||}{DFT} & \multicolumn{3}{c||}{APE$_\mathrm{DFT}^\mathrm{PP/XC}$}\\\hline
												\backslashbox{PP~~}{XC}
												& LDA & PBE & PBEsol & LDA & PBE & PBEsol\\\hline
												LDA & \textbf{159.425} & 111.387 & 128.328 & 0 & 0.965 & 0.364\\
												PBE	& 157.240 &	\textbf{110.321} &	127.513 & \cellcolor{lightgray}1.371 & 0 & 0.273\\
												PBEsol & 157.731 & 111.163 & \textbf{127.862} & 1.063 & 0.763 & 0\\\hline
											\end{tabular}\\[0.5\baselineskip]
											
											\begin{tabular}{|c|c|c||c|c||}\hline
												C$_\mathrm{11}$ & \multicolumn{2}{c||}{DFT} & \multicolumn{2}{c||}{APE$_\mathrm{DFT}^\mathrm{PP/XC}$}\\\hline
												\diagbox{PP~~}{XC}
												& PBE & SCAN & PBE & SCAN\\\hline
												PBE & \textbf{105.489} & 88.918 & 0 & \cellcolor{lightgray}3.248\\
												SCAN & 106.232 & \textbf{91.903} & 0.704 & 0\\\hline
											\end{tabular}\\[1.5\baselineskip]
										}
										\begin{tabular}{|c|c|c|c||c|c|c||c|c|c||c|c||}\hline
											C$_\mathrm{11}$ & \multicolumn{3}{c||}{DFT} & \multicolumn{3}{c||}{APE$_\mathrm{DFT}^\mathrm{PP/XC}$} & C$_\mathrm{11}$ & \multicolumn{2}{c||}{DFT}& \multicolumn{2}{c||}{APE$_\mathrm{DFT}^\mathrm{PP/XC}$}\\\hline
											\backslashbox{PP}{XC} & LDA & PBE & PBEsol & LDA & PBE & PBEsol & \backslashbox{PP}{XC} & PBE & SCAN & PBE & SCAN\\\hline
											LDA & \textbf{159.425} & 111.387 & 128.328 & 0 & 0.965 & 0.364 & PBE & \textbf{105.489} & 88.918 & 0 & \cellcolor{lightgray}3.248\\
											PBE & 157.240 &	\textbf{110.321} &	127.513 & \cellcolor{lightgray}1.371 & 0 & 0.273 & SCAN & 106.232 & \textbf{91.903} & 0.704 & 0\\\cline{8-12}
											PBEsol & 157.731 & 111.163 & \textbf{127.862} & 1.063 & 0.763 & 0\\\cline{1-7}
										\end{tabular}\\[1.5\baselineskip]
										
										\comm{\begin{tabular}{|c|c|c|c||c|c|c||}\hline
												C$_\mathrm{12}$ & \multicolumn{3}{c||}{DFT} & \multicolumn{3}{c||}{APE$_\mathrm{DFT}^\mathrm{PP/XC}$}\\\hline
												\backslashbox{PP~~}{XC}
												& LDA & PBE & PBEsol & LDA & PBE & PBEsol\\\hline
												LDA & \textbf{50.691} & 46.634 & 46.008 & 0 & 1.039 & 0.551\\
												PBE & 49.962 & \textbf{46.154} & 45.383 & \cellcolor{lightgray}1.438 & 0 & 0.816\\
												PBEsol & 50.253 & 46.513 & \textbf{45.756} & 0.864 & 0.776 & 0\\\hline
											\end{tabular}\\[0.5\baselineskip]
											
											\begin{tabular}{|c|c|c||c|c||}\hline
												C$_\mathrm{12}$ & \multicolumn{2}{c||}{DFT} & \multicolumn{2}{c||}{APE$_\mathrm{DFT}^\mathrm{PP/XC}$}\\\hline
												\diagbox{PP~~}{XC}
												& PBE & SCAN & PBE & SCAN\\\hline
												PBE & \textbf{42.306} & 39.017 & 0 & \cellcolor{lightgray}1.574\\
												SCAN & 42.405 & \textbf{39.641} & 0.234 & 0\\\hline
											\end{tabular}\\[1.5\baselineskip]
										}
										\begin{tabular}{|c|c|c|c||c|c|c||c|c|c||c|c||}\hline
											C$_\mathrm{12}$ & \multicolumn{3}{c||}{DFT} & \multicolumn{3}{c||}{APE$_\mathrm{DFT}^\mathrm{PP/XC}$} & C$_\mathrm{12}$ & \multicolumn{2}{c||}{DFT}& \multicolumn{2}{c||}{APE$_\mathrm{DFT}^\mathrm{PP/XC}$}\\\hline
											\backslashbox{PP}{XC} & LDA & PBE & PBEsol & LDA & PBE & PBEsol & \backslashbox{PP}{XC} & PBE & SCAN & PBE & SCAN\\\hline
											LDA & \textbf{50.691} & 46.634 & 46.008 & 0 & 1.039 & 0.551 & PBE & \textbf{42.306} & 39.017 & 0 & \cellcolor{lightgray}1.574\\
											PBE & 49.962 & \textbf{46.154} & 45.383 & \cellcolor{lightgray}1.438 & 0 & 0.816 & SCAN & 42.405 & \textbf{39.641} & 0.234 & 0\\\cline{8-12}
											PBEsol & 50.253 & 46.513 & \textbf{45.756} & 0.864 & 0.776 & 0\\\cline{1-7}
										\end{tabular}\\[1.5\baselineskip]
										
										\comm{\begin{tabular}{|c|c|c|c||c|c|c||}\hline
												C$_\mathrm{44}$ & \multicolumn{3}{c||}{DFT} & \multicolumn{3}{c||}{APE$_\mathrm{DFT}^\mathrm{PP/XC}$}\\\hline
												\backslashbox{PP~~}{XC}
												& LDA & PBE & PBEsol & LDA & PBE & PBEsol\\\hline
												LDA & \textbf{68.824} & 61.691 & 63.175 & 0 & 1.494 & 0.942\\
												PBE & 67.432 & \textbf{60.783} & 62.018 & \cellcolor{lightgray}2.023 & 0 & 0.908\\
												PBEsol & 67.976 & 61.2905 & \textbf{62.586} & 1.231 & 0.835 & 0\\\hline
											\end{tabular}\\[0.5\baselineskip]
											
											\begin{tabular}{|c|c|c||c|c||}\hline
												C$_\mathrm{44}$ & \multicolumn{2}{c||}{DFT} & \multicolumn{2}{c||}{APE$_\mathrm{DFT}^\mathrm{PP/XC}$}\\\hline
												\diagbox{PP~~}{XC}
												& PBE & SCAN & PBE & SCAN\\\hline
												PBE & \textbf{59.379} & 57.897 & 0 & \cellcolor{lightgray}1.235\\
												SCAN & 59.936 & \textbf{58.621} & 0.938 & 0\\\hline
											\end{tabular}\\[1.5\baselineskip]
										}
										\begin{tabular}{|c|c|c|c||c|c|c||c|c|c||c|c||}\hline
											C$_\mathrm{44}$ & \multicolumn{3}{c||}{DFT} & \multicolumn{3}{c||}{APE$_\mathrm{DFT}^\mathrm{PP/XC}$} & C$_\mathrm{44}$ & \multicolumn{2}{c||}{DFT}& \multicolumn{2}{c||}{APE$_\mathrm{DFT}^\mathrm{PP/XC}$}\\\hline
											\backslashbox{PP}{XC} & LDA & PBE & PBEsol & LDA & PBE & PBEsol & \backslashbox{PP}{XC} & PBE & SCAN & PBE & SCAN\\\hline
											LDA & \textbf{68.824} & 61.691 & 63.175 & 0 & 1.494 & 0.942 & PBE & \textbf{59.379} & 57.897 & 0 & \cellcolor{lightgray}1.235\\
											PBE & 67.432 & \textbf{60.783} & 62.018 & \cellcolor{lightgray}2.023 & 0 & 0.908 & SCAN & 59.936 & \textbf{58.621} & 0.938 & 0\\\cline{8-12}
											PBEsol & 67.976 & 61.2905 & \textbf{62.586} & 1.231 & 0.835 & 0\\\cline{1-7}
										\end{tabular}\\[1.5\baselineskip]
										{\begin{tabular}{|c|c|c|c||c|c|c||c|c|c||c|c||}\hline
												\textbf{$\gamma$} & \multicolumn{3}{c||}{DFT} & \multicolumn{3}{c||}{APE$_\mathrm{DFT}^\mathrm{PP/XC}$} & \textbf{$\gamma$} & \multicolumn{2}{c||}{DFT}& \multicolumn{2}{c||}{APE$_\mathrm{DFT}^\mathrm{PP/XC}$}\\\hline
												\backslashbox{PP}{XC} & LDA & PBE & PBEsol & LDA & PBE & PBEsol & \backslashbox{PP}{XC} & PBE & SCAN & PBE & SCAN\\\hline
												LDA & \textbf{0.502} & 0.305 & 0.362 & 0 & \cellcolor{lightgray}3.400 & 1.784 & PBE & \textbf{0.364} & 0.351 & 0 & \cellcolor{lightgray}0.392\\
												PBE & 0.512 & \textbf{0.316} & 0.371 & 1.921 & 0 & 0.623 & SCAN & 0.365 & \textbf{0.350} & 0.350 & 0\\\cline{8-12}
												PBEsol & 0.511 & 0.313 & \textbf{0.369} & 1.765 & 0.934 & 0\\\cline{1-7}
											\end{tabular}\\[1.5\baselineskip]}
										
										\begin{tabular}{|c|c|c|c||c|c|c||c|c|c||}\hline
											\textbf{B}$^\mathrm{EXP}$=76.3 & \multicolumn{3}{c||}{DFT} & \multicolumn{3}{c||}{APE$_\mathrm{DFT}^\mathrm{PP/XC}$} & \multicolumn{3}{c||}{APE$_\mathrm{EXP}^\mathrm{PP/XC}$}\\\hline
											\backslashbox{PP~~}{XC}
											& LDA & PBE & PBEsol & LDA & PBE & PBEsol & LDA & PBE & PBEsol\\\hline
											LDA & \textbf{86.936} &	68.218 & 73.448 & 0 & 0.999 & 0.442 & \cellcolor{lightgray}13.939 & 10.592 & 3.738\\
											PBE & 85.721 & \textbf{67.543} & 72.760 & \cellcolor{lightgray}1.397 & 0 & 0.500 & 12.348 & 11.476 & 4.640\\
											PBEsol & 86.079 & 68.063 & \textbf{73.125} & 0.985 & 0.769 & 0 & 12.817 & 10.796 & 4.161\\\hline
										\end{tabular}\\[0.5\baselineskip]
										
										\begin{tabular}{|c|c|c||c|c||c|c||}\hline
											\textbf{B}$^\mathrm{EXP}$=76.3 & \multicolumn{2}{c||}{DFT} & \multicolumn{2}{c||}{APE$_\mathrm{DFT}^\mathrm{PP/XC}$} & \multicolumn{2}{c||}{APE$_\mathrm{EXP}^\mathrm{PP/XC}$}\\\hline
											\diagbox{PP~~}{XC}
											& PBE & SCAN & PBE & SCAN & PBE & SCAN\\\hline
											PBE & \textbf{63.367} &	55.651 & 0 & \cellcolor{lightgray}2.473 & 16.950 & \cellcolor{lightgray}27.063\\
											SCAN & 63.681 & \textbf{57.062} & 0.495 & 0 & 16.539 & 25.214\\\hline
										\end{tabular}\\[1.5\baselineskip]
										
										\begin{tabular}{|c|c|c|c||c|c|c||}\hline
											\textbf{$\Delta_i$} & \multicolumn{3}{c||}{\textbf{$\Delta_i$}(a,b)} & \multicolumn{3}{c||}{\textbf{$\Delta_i$}(a,b)}\\\hline
											\backslashbox{PP~~}{XC} & LDA & PBE & PBEsol & \backslashbox{PP~~}{XC} & PBE & SCAN\\\hline
											LDA & 0 & 1.168 & 0.758 & PBE & 0 & 0.261\\
											PBE & \cellcolor{lightgray}1.386 & 0 & 0.452 & SCAN & \cellcolor{lightgray}0.337 & 0 \\\cline{5-7}
											PBEsol & 0.852 & 0.440 & 0 \\\cline{1-4}
										\end{tabular}
									\end{table}

\end{suppinfo}
}

\end{document}